\documentclass[preprint]{aastex}
\usepackage{graphicx}
\usepackage{amsmath}
\usepackage{color} 
\usepackage{epsfig}
\usepackage{subfigure}
\usepackage{txfonts}

\newcommand{\Ms}{M_{\odot}}

\newcommand{\Mj}{M_{J}}

\newcommand{\tot}{{\rm tot}}

\graphicspath{{figures/}}

\makeatother
\shorttitle{Hierarchical secular coupling of two planets}
\shortauthors{Teyssandier et al.}
\makeatother

\begin{document}
\title{Extreme orbital evolution from hierarchical secular coupling of two giant planets}
\author{Jean Teyssandier\altaffilmark{1,\dagger}, Smadar Naoz\altaffilmark{2,3}, Ian Lizarraga\altaffilmark{4}, Frederic A. Rasio\altaffilmark{3,5}}
\altaffiltext{1}{Institut d'Astrophysique de Paris, UPMC Paris 06, CNRS, UMR7095, 98 bis bd Arago, F-75014, Paris, France}
\altaffiltext{2}{Harvard Smithsonian Center for Astrophysics, Institute for Theory and Computation, 60 Garden St., Cambridge, MA 02138}
\altaffiltext{3}{Center for Interdisciplinary Exploration and Research in Astrophysics (CIERA), Northwestern University, Evanston, IL 60208, USA}
\altaffiltext{4}{Center for Applied Mathematics, Cornell University, Ithaca, NY 14853-3801}
\altaffiltext{5}{Department of Physics and Astronomy, Northwestern University}
\altaffiltext{$\dagger$}{Email: teyssand@iap.fr}


\begin{abstract}

Observations of exoplanets over the last two decades have revealed a new class of Jupiter-size  planets with orbital periods of a few days, the so-called ``hot Jupiters''. Recent measurements using the Rossiter--McLaughlin effect have shown that many ($\sim 50\%$) of these planets are misaligned; furthermore, some ($\sim 15\%$) are even retrograde with respect to the stellar spin axis. Motivated by these observations, we explore the possibility of forming retrograde orbits in hierarchical triple configurations consisting of a star--planet inner pair with another giant planet, or brown dwarf, in a much wider orbit. Recently \citet{SN11a} showed that in such a system, the inner planet's orbit can flip back and forth from prograde to retrograde, and can also reach extremely high eccentricities. Here we map a significant part of the parameter space of dynamical outcomes for these systems. We derive strong constraints on the orbital configurations for the outer perturber (the tertiary) that could lead to the formation of hot Jupiters with misaligned or retrograde orbits. We focus only on the secular evolution, neglecting other dynamical effects such as mean-motion resonances, as well as all dissipative forces. For example, with an inner Jupiter-like planet initially on a nearly circular orbit at  $5\,$AU, we show that a misaligned hot Jupiter is likely to be formed in the presence of a more massive planetary companion ($> 2\Mj$) within $\sim140\,$AU of the inner system, with mutual inclination  $>50^\circ$ and eccentricity above $\sim0.25$. This is in striking contrast to the test-particle approximation, where an almost perpendicular configuration can still cause large eccentricity excitations, but flips of an inner Jupiter-like planet are much less likely to occur. The constraints we derive  can be used to guide future observations, and, in particular, searches for more distant companions in systems containing a hot Jupiter.
\end{abstract}



\section{Introduction}
\label{sec:intro}
To date, about 800 exoplanets have been detected. This number is growing sharply, with more and more planet candidates from the \textit{Kepler} catalogue being confirmed \citep[there are currently about 3400 unconfirmed candidates, with an overall false-positive rate expected to be $9.4\pm0.9\% $ according to ][]{Fressin13}. Some of the earliest detections led to the surprising discovery of a new class of Jupiter-like planets in very close proximity to their host star \citep{MQ95}, the so-called ``hot Jupiters'' (hereafter HJ). {\it In situ\/} formation at such short distances (just a few stellar radii) from the parent star seems very unlikely. A popular explanation for the presence of a giant gas planet so close to the star  is planetary migration, associated with viscous evolution of protoplanetary disk \citep{Lin_Papaloizou,Mass+03}. This migration should result in orbits with low eccentricities and inclinations \citep[but see ][]{Lai+10,Thies+11}. However, it was shown that other dynamical mechanisms such as planet--planet scattering \citep{RasioFord,TP2002,Nagasawa,Nag+11,CFMR,WuLithwick2011,BN,Boley+12} and secular evolution \citep{Hol+97,Wu2003,Takeda05,Wu2007,KCTF,Takeda,SN11a,Correia2011,Naoz+12bin,KP12,SN13} also play an important role in the formation of HJs.

The Rossiter--McLaughlin effect \citep{Rossiter,Mclaughlin,RML} has enabled measurement of the sky-projected angle between the orbits of several HJs and the spins of their host stars. Surprisingly, about half of these planets are observed to be misaligned and some (about $25\%$) are even in retrograde orbits with respect to the spin axis of the host star \citep[e.g.,][]{Triaud,Albrecht+12,Brown2012}. These observations suggest that the classical disk migration model is not the only channel to form HJs.

The \textit{Kepler} mission has so far revealed the existence of about 2300 planet candidates, and the number of false positives among this sample is expected to be small \citep{Batalha2012,Morton12}. A recent analysis by \citet{Steffen2012} showed that most HJs from the \textit{Kepler} data appear to have no nearby, coplanar companions (within a period ratio of a few); however, planetary companions at larger separations and large inclinations cannot be excluded (especially since an outer companion with an orbital period ratio of $\sim$ 10  and a $60^{\circ}$ mutual inclination would have a detection likelihood of less than $5\%$  by transit methods). 

Recent developments in direct imaging provide a powerful tool to detect a class of planets that cannot be observed via  radial velocity or transit methods: massive planets with large angular separation (i.e., within orbits of tens of astronomical units). For example, \citet{Lafreniere2008} \citep[see also][]{Lafreniere2010} found evidence of the first directly imaged planet (of  $8\, \Mj$\footnote{Hereafter we denote by $\Mj$ the mass of Jupiter} and a separation of $\sim330$~AU) around a young Sun-like star. Shortly after this observation, \citet{Marois08} announced the discovery of a system of three planets orbiting at several tens of AU of the HR 8799 star, with masses ranging from 5 to 13 $\Mj$ \citep[see also][for the discovery of a fourth inner planet]{Marois10,Skemer12}. More recently, two additional planets (with masses of about $4\,  \Mj$) were discovered through direct imaging at projected distances of few tens of AU from their host stars \citep{Rameau,Kuzuhara13}. Planets at such distances could have formed in situ through  gravitational instabilities in a massive protoplanetary disk \citep{Durisen}, or could have been brought there through outward disk-driven migration of planets formed at distances of about 10 AU \citep{Crida09}. Alternatively  it is possible that these planets have migrated there as the result of strong gravitational interactions with (at least) another planet in the system, suggesting a multiple-planet system. Therefore, populations of planets on both close and wide orbits might coexist in planetary systems. These populations in principle could have a large range of eccentricities and inclinations, because of their dynamical history.
The direct imaging method is more effective in young systems, for which the planet at large separation still has an important thermal emission, making it easier to observe. Unfortunately, this limitation affects the possibility of detecting any close-in planets (since the star is still very active). Astrometry is another promising method for detecting planets on wide orbits. The efficiency of this method increases with both the orbital separation and the mass of the planet. Therefore a mission such as \textit{Gaia} could give new insights in the detection of massive planets with orbital periods of several years \citep[see, e.g.,][and references therein]{Sozzetti13}, with access to a wide range of orbital parameters. In addition, a distant planetary perturber causes a long-term linear trend in the radial velocity curve of its host star, which could appear in long-term radial velocity surveys \citep[see, e.g.,][]{Crepp12}. However this trend does not have a significant effect on the systems we study; we quantify this effect in our results.

Different theoretical models have been proposed to explain the presence of HJs and the observed misalignments in particular. Some studies proposed that dynamical gravitational scattering in multiplanet systems can lead to large eccentricities and misaligned HJs \citep{RasioFord,TP2002,Nagasawa,Nag+11,CFMR,WuLithwick2011,BN,Boley+12}. Other studies invoke secular effects  (i.e., interactions on timescales that are long compared to the orbital period) by stellar or planet companions in the dynamical evolution of planetary systems in the framework of triple systems  \citep{Hol+97,Wu2003,Takeda05,Wu2007,KCTF,Takeda,SN11a,Correia2011,Naoz+12bin,KP12,SN13}. Furthermore, different models suggest that misalignment can be caused by magnetic interactions between the protoplanetary disk and the parent star \citep{Lai+10} or dynamical interactions with another star that would tilt the disk's axis \citep{Thies+11}. Therefore the planets formed in such disks would be naturally misaligned. \citet{Chen2013} also showed that a combination of disk--planet secular interactions with subsequent Kozai oscillations between the two planets can produce misaligned HJs. In addition, simulations by \citet{TTP12}, \citet{Xiang13} and \citet{Bitsch13} also showed that if planets on inclined orbits cohabited with a disk, massive planets were likely to align with the disk, as inclination damping occurs on a timescale shorter than the lifetime of the disk, whereas less massive planets would remain on inclined and eccentric orbits because of Kozai-like excitations emerging from interactions with the disk. 

Here we study the parameter space of a planetary perturber in the framework of triple-body dynamics. For arbitrary inclinations and eccentricities, long-term stability requires the system to be hierarchical. Therefore, the system must consist of an ``inner'' binary (stellar mass $m_0$ and Jupiter mass $m_1$) in a nearly-Keplerian orbit with semi major axis (SMA) $a_1$, and an ``outer'' binary in which $m_2$ orbits the center of mass of the inner binary, with SMA $a_2 \gg a_1$. Another condition for stability is that the eccentricity of the outer orbit, $e_2$, cannot be too large so that $m_2$ does not make close approaches to the inner binary orbit. In such systems a high mutual inclination between $m_2$ and the $(m_0,m_1)$ system can produce large-amplitude oscillations of the eccentricity and inclination; this is the so-called Kozai--Lidov mechanism \citep{Kozai,Lidov}.

\citet{Kozai} studied the effects of Jupiter's gravitational perturbation on an inclined asteroid in our own solar system  using Hamiltonian perturbation theory. In this influential work,  Jupiter was assumed to be on a circular orbit, thus the  massless asteroid moved in an axisymmetric gravitational potential. The immediate consequence is that the projection of the inner orbit's angular momentum along the total angular momentum is conserved during the evolution. In fact, at the lowest order of approximation in the ratio of semi-major axes, $\alpha= a_1/a_2$, (called the ``quadrupole'' approximation) in the {\it test particle} case (i.e., one of the objects in the inner binary is  massless), the component of the inner orbit's angular momentum along the total is conserved even if the outer orbit is not circular \citep[e.g.,][]{LZ74}. Recently, \citet{SN11a,SN13} showed that these approximations are not appropriate for many systems, particularly in the presence of a (minimally) eccentric outer orbit when the next-order perturbations (octupole) are taken into account, or if the test particle approximation for the inner body is relaxed (at quadrupole or octupole order). As a consequence, the relevant component of the angular momentum is no longer conserved.  The lack of conservation of the inner orbit's angular momentum component allows the orbit to reach extremely high eccentricities and can even ``flip'' the orbit from prograde to retrograde with respect to the total angular momentum.

\citet{SN11a} considered the secular evolution of a triple system consisting of an inner binary containing a star and a Jupiter-like planet separated by several AU, orbited by a distant Jupiter-like planet or brown dwarf companion. Perturbations from the outer body can drive Kozai-like cycles in the inner binary, which, when planet--star tidal effects are incorporated, can lead to the capture of the inner planet. This leads to a close, highly inclined or even retrograde orbit, similar to the orbits of the observed misaligned HJs. 

Here we explore the orbital parameter space of a triple-body hierarchical system in the point mass limit (i.e., neglecting tidal dissipation). We focus on planetary systems, where the perturbing object is either a planet or a brown dwarf, but as we will show the system can be scaled to different masses. We show that going beyond the test particle approximation yields qualitatively different results. We map the parameter space of the outer orbit in terms of mass, separation, eccentricity and inclination, and seek the best configurations that would produce retrograde orbits. Thus, we predict the properties of the planet perturber that causes the eccentric Kozai--Lidov evolution. The eccentricity of the inner planet grows large enough to trigger tidal circularization around the host star and could eventually form misaligned HJs. We do not study this process in this paper but give constraints on the perturber that can trigger and cause this behavior. This can help guide future observational programs.

This paper is organized as follows: in Section \ref{sec:kozai} we review the main features of the eccentric Kozai--Lidov mechanism. In Section \ref{sec:res} we present the results of our numerical study: in Section \ref{sec:ratio} we map the complete space of parameters, finding the best configurations that could allow the orbit to flip in a retrograde motion and in Section \ref{sec:cumul_dist} we look in closer detail at the inner eccentricity distribution. In Section \ref{sec:MC} we run a set of Monte Carlo simulations in order to study precisely the outcome of two representative cases. Finally we discuss these results in section \ref{sec:discuss}.


\section{Secular perturbations with an eccentric perturber}
\label{sec:kozai}

\subsection{The Eccentric Kozai-Lidov mechanism}

Throughout the paper we consider the evolution of two planets of mass $m_1$ and $m_2$ orbiting a central star of mass $m_0$. The subscript 1 refers to the inner orbit (consisting of the central mass $m_0$ and the inner planet $m_1$), and the subscript 2 refers to the outer orbit (consisting of the inner orbit's center of mass and the $m_2$ planet). For $k = 1,2$, we denote by $a_k$, $e_k$ and $i_k$ the SMA, eccentricity and inclination of the inner (1) and outer (2) orbits respectively. Throughout the paper we refer to the inclination angle of the inner (outer) orbit with respect to the total angular momentum, i.e., $i_1$ ($i_2$), and to the mutual inclination between the two orbits, which  is simply $i_\tot=i_1+i_2$. 

Relaxing the test particle approximation, \citet{SN11a,SN13} showed that even in the quadrupole level of approximation one finds deviations  from the ``classical'' quadrupole level Kozai evolution.  Specifically, the inclination can oscillate around $90^\circ$  \citep[e.g.,][]{SN13}, where in the ``classical'' Kozai mechanism the quantity $\sqrt{1-e_1^2}\cos i_1$ is constant, thus forbidding flips from prograde to retrograde orbits.
 
The ``classical'' Kozai mechanism is valid for the lowest (quadrupole) order of approximation  (if applicable), and if one of the inner orbit members is a test particle. We refer to this limit as the test particle quadrupole (TPQ) approximation. Here we relax the TPQ approximation. In addition we focus on eccentric perturbers, which emphasize the need for the octupole level of approximation \citep[e.g.,][]{SN11a,SN13}. 
In all of our runs we use the Bulirsch--Stoer method in order to numerically solve the octupole-level secular equations following \citet{SN13}, including first-order post-Newtonian relativistic precession of the inner and outer orbits  \citep[e.g.,][ note that the interaction term presented there does not affect our results here]{Naoz+13GR}. We compare our results to the octupole-level test particle approximation \citep{LN11,BK}.

\begin{figure}[t!]
      \centering \includegraphics[scale=1]{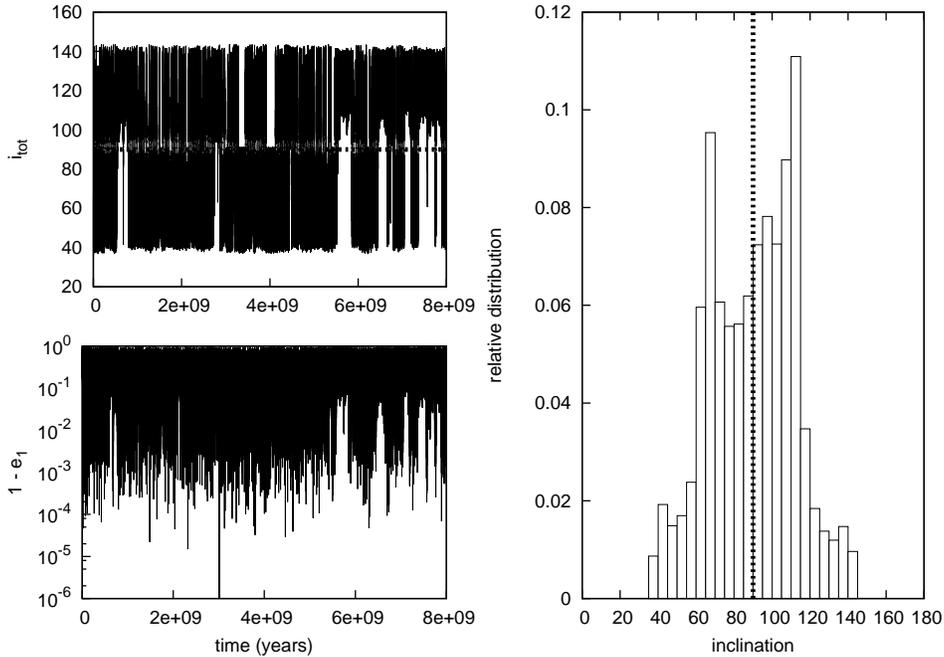}
      \caption{Time evolution of the nominal example. \textit{Left}: evolution of the mutual inclination (top) and eccentricity  of the inner planet (bottom, as $1-e_1$ in log scale) for a two-planet system. The horizontal dashed line shows the separation between prograde and retrograde orbits at $90^{\circ}$. \textit{Right}: relative distribution of the mutual inclination during the integration time. The system is the following: a $1\, \Ms$ star with an inner planet of $1\, \Mj$ on an initially circular orbit at 5 AU, and an outer planet of $6\, \Mj$ at 61 AU with an eccentricity of 0.5. The two orbits are initially separated by $65^{\circ}$. The vertical dashed line shows the separation between prograde and retrograde orbits at $90^{\circ}$.} 
      \label{distrib}
\end{figure}

In the octupole level of approximation, the inner orbit's eccentricity can reach very high values, which we map below (see Section \ref{sec:cumul_dist}). In addition the inner orbit's inclination can flip its orientation from prograde ($i_\tot<90^\circ$), with respect to the total angular momentum, to retrograde  ($i_\tot>90^\circ$).  We refer to this process as the eccentric Kozai--Lidov mechanism \citep[hereafter EKL, following the notation of][]{Naoz+12bin}. In  Figure \ref{distrib} we show an example for the time evolution of a system that is influenced by the EKL mechanism. The system is set initially with the following parameters: $m_0=1\, \Ms$, $m_1=1\, \Mj$, $m_2=6\, \Mj$, $a_1=5$~AU, $a_2=61$~AU, $e_1=0.01$, $e_2=0.5$, $i_\tot=65^\circ$, with the arguments of pericenters set to be $g_1=g_2=0^\circ$. The longitudes of ascending nodes are set by the relation $h_2-h_1=180^\circ$, \citep[see][]{SN13}. We choose these parameters as our nominal example and we will often compare our result to it. As can be seen in Figure \ref{distrib} the mutual inclination keeps flipping from a prograde orbit ($i_\tot < 90^\circ$) to a retrograde one ($i_\tot > 90^\circ$). However those flips are not regularly spaced in time; at every flip the time spent on a prograde or retrograde orbit is not the same. Nevertheless we can see that on average, over the total integration time, the inclination is roughly equally distributed between prograde and retrograde orbits. In addition we note that the eccentricity is mainly distributed between 0 and 0.9 (precisely, 89\% of the integration time is spent between these two values in this simulation), but also reaches very high values (up to 0.9999) which, of course, would not make any sense in a system where tidal friction would take place (we quantify the eccentricity distribution in Section \ref{sec:MC}). In the right panel of Figure \ref{distrib} we show the distribution of the mutual inclination over the integration time. We also quantify the inclination distribution in Section \ref{sec:MC}. For this specific example the distribution shows two peaks located at the initial angle and its symmetric with regard to $90^{\circ}$. Here the inclination tends to be equally distributed between prograde and retrograde orbits. Also, because the system started initially with zero eccentricity the nominal Kozai critical angles ($40^{\circ}$ and $140^{\circ}$) are limiting the system.

\subsection{Timescales}
\label{sec:ts}

\begin{figure}[t!]
      \centering \includegraphics[scale=1]{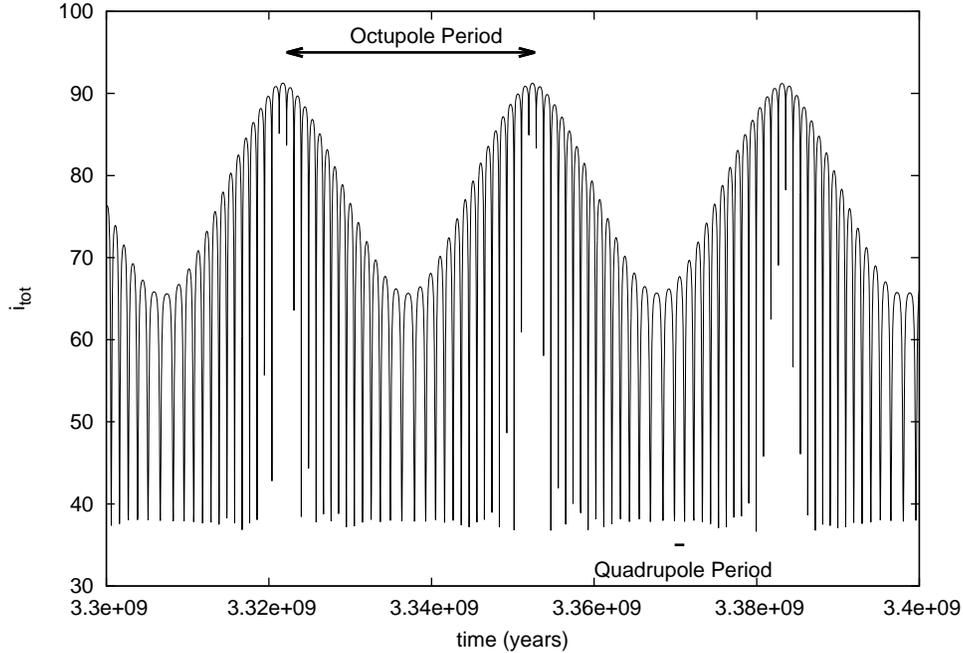}
      \caption{close-up of the time evolution of the inclination of Figure \ref{distrib}. Two periods appear: the Kozai oscillations due to the quadrupole term, and the oscillation of the octupole envelope.} 
      \label{zoomdist}
\end{figure}

In Figure \ref{zoomdist} we show a close- up of Figure \ref{distrib}, where two distinct timescales appear. Both of them can be associated with a term of the Hamiltonian expansion of the hierarchical three-body problem. The shorter period arises from the quadrupole term, and the longer one arises from the octupole term. These timescales can be estimated by Equations (\ref{eq:tquad}) and (\ref{eq:toct}) respectively, where $k^2$ is the gravitational constant \citep[see, e.g.,][with a modification for the octupole timescale, taking into account the inclination]{Naoz+13GR}:

\begin{equation}\label{eq:tquad}
t_{\rm quad}\sim \frac{2\pi a_2^3 (1-e_2^2)^{3/2}\sqrt{m_0+m_1}}{ a_1^{3/2} m_2 k}  \ ,
\end{equation}

\begin{equation}
\label{eq:toct}
t_{\rm oct}\sim  2\pi \frac{4}{15} \frac{a_2^4 (1-e_2^2)^{5/2} \sqrt{1-e_1^2} (m_0+m_1)^{3/2}}{a_1^{5/2}e_2 k |m_0-m_1| m_2 } \frac{1}{\frac{G_1}{G_2}+\cos{i_{\rm tot}}}\ ,
\end{equation}
where  $G_1$ and $G_2$ are the magnitudes of the angular momenta of each orbits, and are given by 
\begin{align}
\label{eq:G1G2}
G_1 &= \frac{m_0 m_1}{m_0+m_1}\sqrt{k^2(m_0+m_1)a_1(1-e_1^2)},\\
G_2 &= \frac{m_2(m_0+m_1)}{m_0+m_1+m_2}\sqrt{k^2(m_0+m_1+m_2)a_2(1-e_2^2)}.
\end{align}

From Equation (\ref{eq:toct}) we see that the octupole timescale increases sharply toward high inclinations. Therefore, systems with very high mutual inclinations (close to polar configurations) are less likely to flip from prograde to retrograde, because the octupole effects take place on a longer timescale than for moderately inclined systems \citep[we refer the reader to][for further discussions on the octupole timescale]{Li2013}. From Figure \ref{zoomdist} we see that in a system that regularly flips from prograde to retrograde configurations, the octupole timescale is of the order of $10^{7}$\, yr. Hence we can expect that for similar initial configurations, this will be the typical timescale for a flip.

We also consider the timescale arising from the  relativistic precession of the pericenter of the inner orbit \citep[see, e.g,][]{Naoz+13GR}, 
\begin{equation}
\label{eq:tgr}
t_{\rm 1PN,1} \sim  2\pi \frac{a_1^{5/2} c^2 (1-e_1^2) }{ 3 k^3 (m_0+m_1)^{3/2}} \ ,
\end{equation}
where $c$ is the speed of light. The same precession timescale can be derived for the outer planet, replacing the subscript 1 by 2, but is negligible because of the large SMA of the outer orbit. If the first-order post-Newtonian (hereafter 1PN) timescale is smaller than the timescale associated with the octupole term, then the latter can be suppressed, leading to no orbital flips of the planet. This will be studied in greater detail in section \ref{sec:ratio}. In addition, the EKL mechanism will be completely suppressed if the 1PN timescale is smaller than the timescale associated with the quadrupole term.

\section{Systematic survey of the parameter space}
\label{sec:res}

In the following section we present numerical results describing the effects of mass, SMA ratio, mutual inclination and the outer orbit's eccentricity. For some specific values of these parameters, the system might be in an unstable configuration. We use the \citet{Mardling+01} stability criterion, which defines a stable three-body system as one that obeys
\begin{equation}
\label{eq:Mar}
\frac{a_2}{a_1}> 2.8(1+q_m)^{2/5}\frac{(1+e_2)^{2/5}}{(1-e_2)^{6/5}}\left(1-0.3\frac{i_\tot}{180}\right ) \ ,
\end{equation}
where $q_m=m_2/(m_0+m_1)$ and $i_\tot$ is in degrees. When necessary, we will clearly indicate which region of the parameter space is likely to be unstable. We can already note that the systems are almost always stable for the parameters we have chosen, especially because $q_m$ is very small in the case of two planets.

The integration time in all our simulations was 8 Gyr. It is important to emphasize that a lower integration time affects the results considerably. In Appendix \ref{sec:conv} we describe our convergence test, which clearly shows that only integration times greater than $5000 t_{\rm quad}$ converge, where $t_{\rm quad}$ is the typical timescale for quadrupole oscillations, and is given by Equation (\ref{eq:tquad}). 

\subsection{Likelihood of flipping the orbit}
\label{sec:ratio}

\begin{deluxetable}{cccccccc}
\label{tab:runs}
\tablewidth{330pt}
\tablecaption{Initial conditions of Figures \ref{map_m2m3_ratio}--\ref{map_ie2_ratio}}
\tablenum{1}
\tablehead{\colhead{Figure}  & \colhead{$m_1$} & \colhead{$m_2$} & \colhead{$a_1$} & \colhead{$a_2$} & \colhead{$e_1$} & \colhead{$e_2$} & \colhead{$i_\tot$}  \\ 
\colhead{} & \colhead{($\Mj$)} & \colhead{($\Mj$)} & \colhead{(AU)} & \colhead{(AU)} & \colhead{} & \colhead{} & \colhead{(deg)} } 
\startdata
\ref{map_m2m3_ratio} & 1-9 & 1-30 & 5 & 61 & 0.01 & 0.5 & 65 \\
\ref{map_a1a2_ratio} & 1 & 6 & 2-20 & 10-250 & 0.01 & 0.5 & 65 \\
\ref{map_a2m3_ratio} & 1 & 1-10 & 5 & 51-201 & 0.01 & 0.5 & 65 \\
\ref{map_e2m3_ratio} & 1 & 1-10 & 5 & 61 & 0.01 & 0.1-0.7 & 65 \\
\ref{map_a2e2_ratio} & 1 & 6 & 5 & 51-201 & 0.01 & 0.1-0.8 & 65 \\
\ref{map_im3_ratio}  & 1 & 1-10 & 5 & 61 & 0.01 & 0.5 & 35-90 \\
\ref{map_a2i_ratio}  & 1 & 6 & 5 & 51-201 & 0.01 & 0.5 & 35-90 \\
\ref{map_ie2_ratio}  & 1 & 6 & 5 & 61 & 0.01 & 0.1-0.7 & 35-90 \\
\enddata
\tablecomments{Initial conditions for Figures \ref{map_m2m3_ratio} to \ref{map_ie2_ratio}. For all these runs, we took the arguments of pericenters to be initially $g_1=g_2=0^{\circ}$.}
\end{deluxetable}

\begin{figure}[h!]
   \begin{minipage}[t]{0.40\linewidth}
      \centering \includegraphics[scale=0.65]{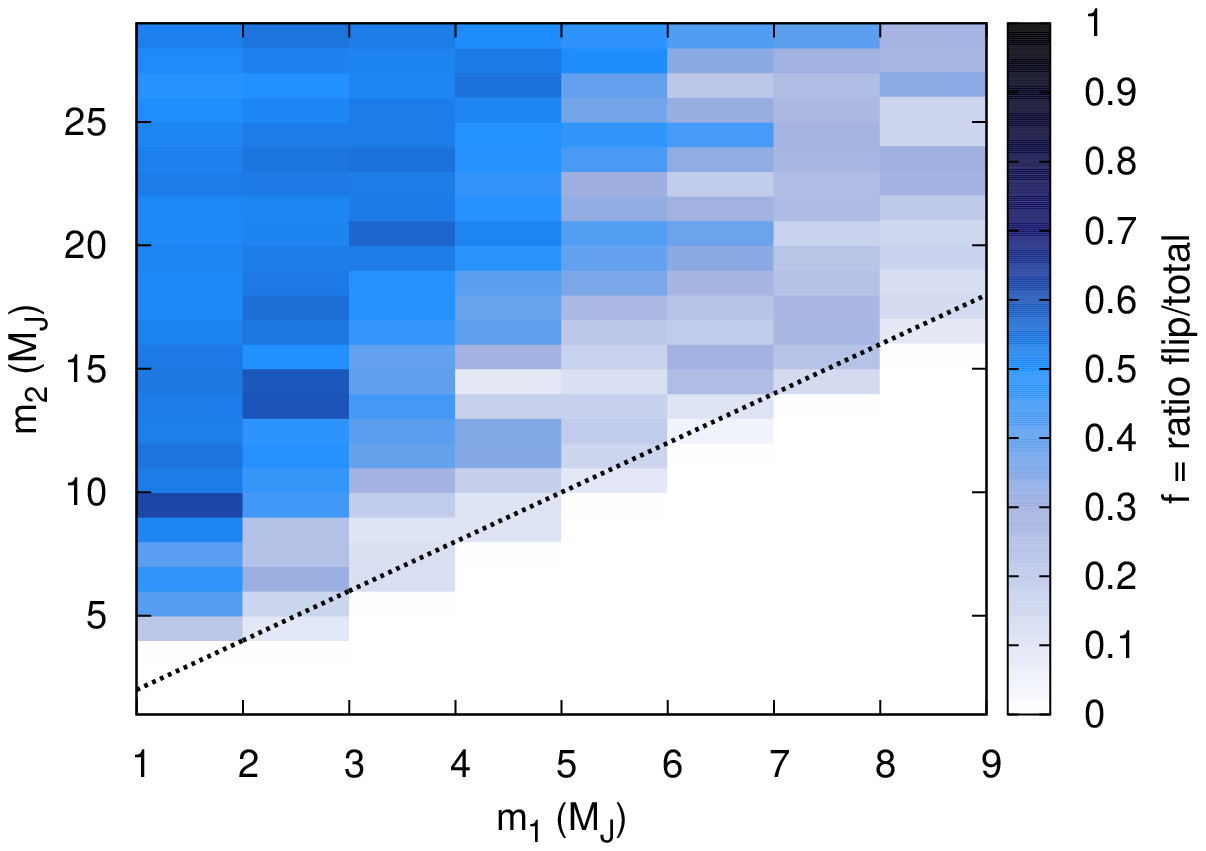}
      \caption{Constant parameters are $a_1=5$~AU, $a_2=61$~AU, $e_{2}=0.5$ and $i_\tot=65^{\circ}$. The black dashed line represent the $m_2=2m_1$ function. When $q=m_1/m_2<0.5$ inner orbits start going retrograde, but only for $q\lesssim 0.3$ do they start to converge to $f\simeq 0.5$.} 
      \label{map_m2m3_ratio}
   \end{minipage}\hfill
   \begin{minipage}[t]{0.48\linewidth}
      \centering \includegraphics[scale=0.65]{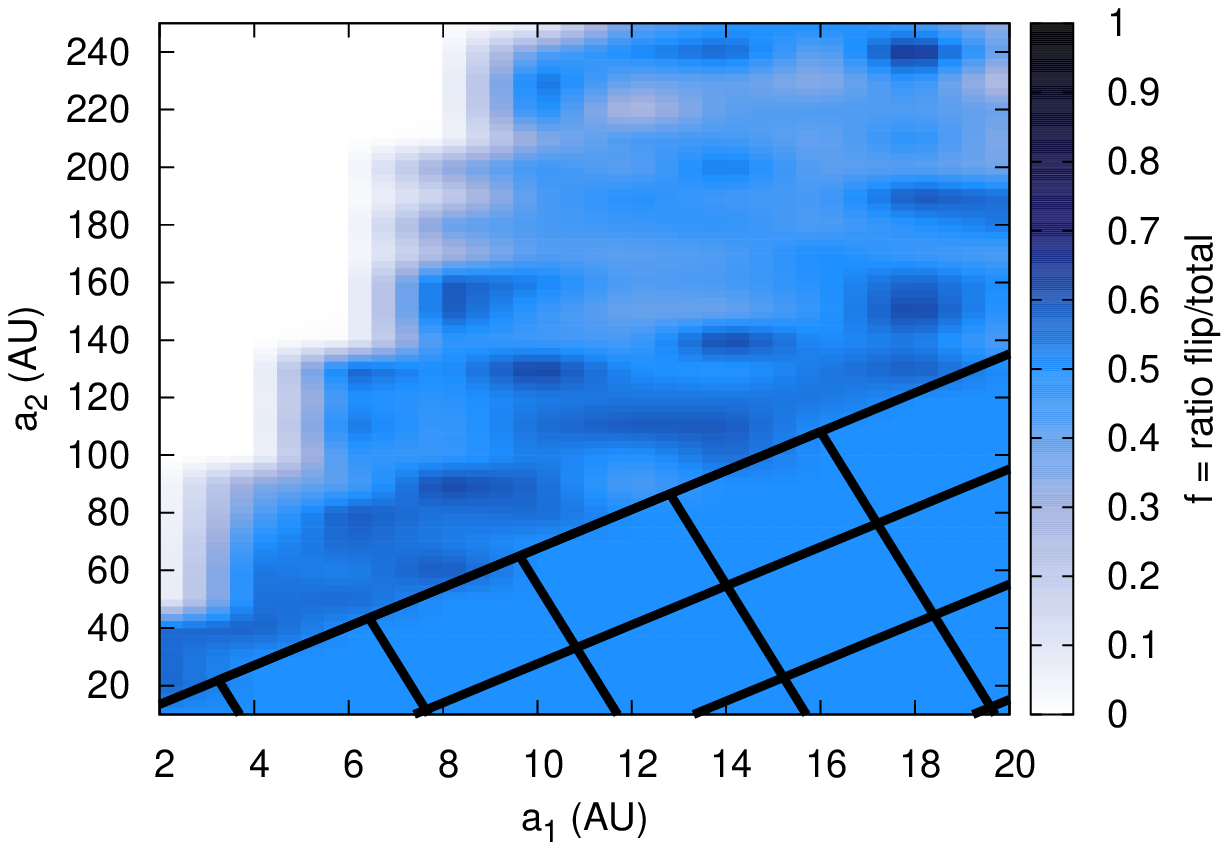}
      \caption{Constant parameters are  $m_1=1\, \Mj$, $m_2=6\, \Mj$, $e_{2}=0.5$ and $i_\tot=65^{\circ}$. The lower right black dashed region denotes the region where orbits are likely to be unstable (see Section 2). Large orbital separations (roughly $a_1/a_2 < 1/25$ for this set of initial conditions) lead to no formation of retrograde orbits ($f=0$).}
      \label{map_a1a2_ratio}
   \end{minipage}\hfill   
\end{figure}

In order to estimate the likelihood of this orbital flip, we compute the  time spent in a retrograde motion ($i_\tot\geq 90^\circ$) over the total integration time ($t_{\tot}$). We define a new dimensionless parameter $f$ by 
\begin{equation}
f=\frac{t(i_\tot \geq 90^\circ)}{t_{\tot}} \ ,
\label{eq:f}
\end{equation}
and we map this variable over the parameter space.  For example a system that never flips from prograde to retrograde has $f=0$, and a system that spends exactly half of its time on a retrograde orbit has $f=0.5$.
We consider our nominal example and systematically vary two parameters in each set of runs (see Table \ref{tab:runs} for a summary of all the parameters). 
We plot $f$ as a function of two of these parameters.  Results are displayed in Figures \ref{map_m2m3_ratio}--\ref{map_ie2_ratio}. For each plot, the initial settings are given in the caption of the figure. In Appendix \ref{sec:max} we map the same numerical experiments as a function of the maximum inclination reached during the integration.

\begin{figure}[h!]
   \begin{minipage}[t]{0.40\textwidth}
      \centering \includegraphics[scale=0.65]{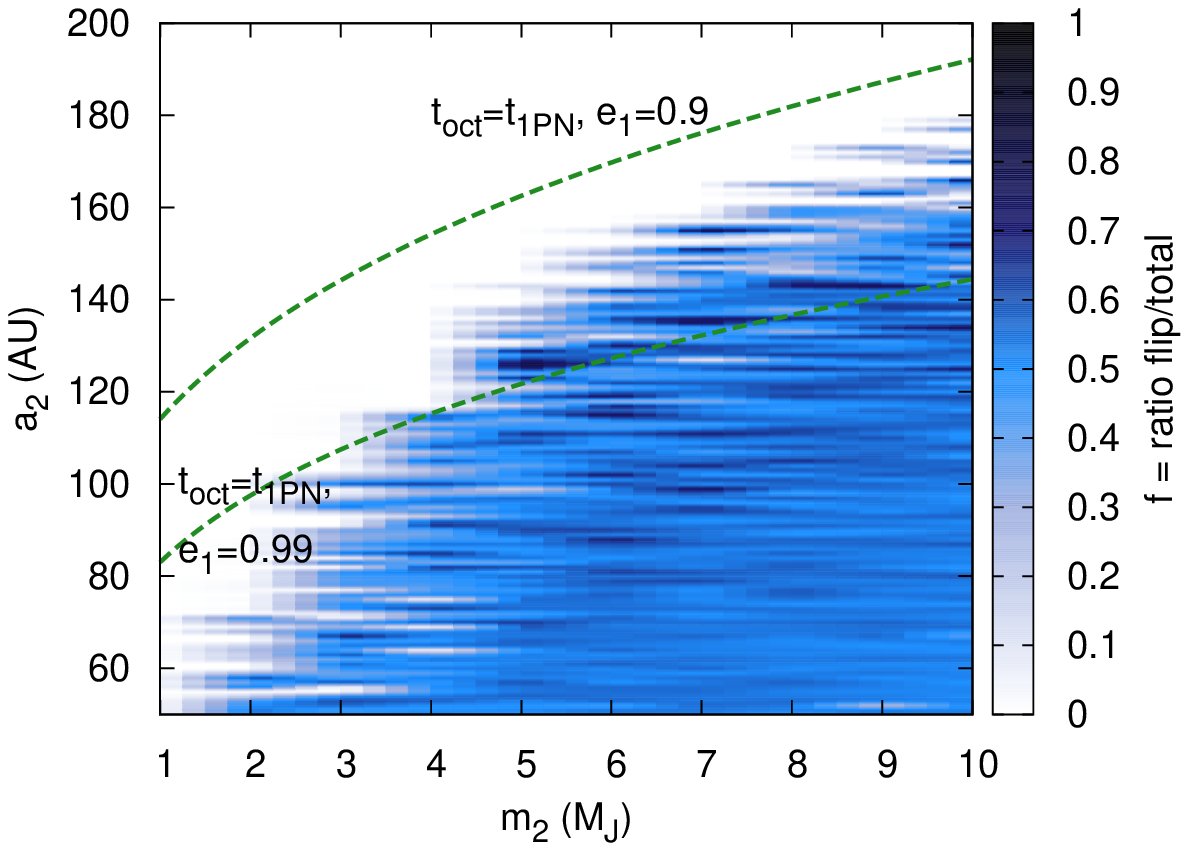}
      \caption{Constant parameters are $m_1=1\, \Mj$, $a_1=5$~AU, $e_2=0.6$ and $i_\tot=65^{\circ}$. The green dashed lines give the approximate location where the octupole timescale is equal to the 1PN timescale. The top one is for $e_1=0.9$ and the bottom one for $e_1=0.99$. Systems on the left-hand side of these lines have a 1PN precession time shorter than the octupole time. A close, massive perturber (i.e., a strong perturbative potential) induces a longer time in retrograde orbits.} 
      \label{map_a2m3_ratio}
   \end{minipage}\hfill
   \begin{minipage}[t]{0.48\linewidth}
      \centering \includegraphics[scale=0.65]{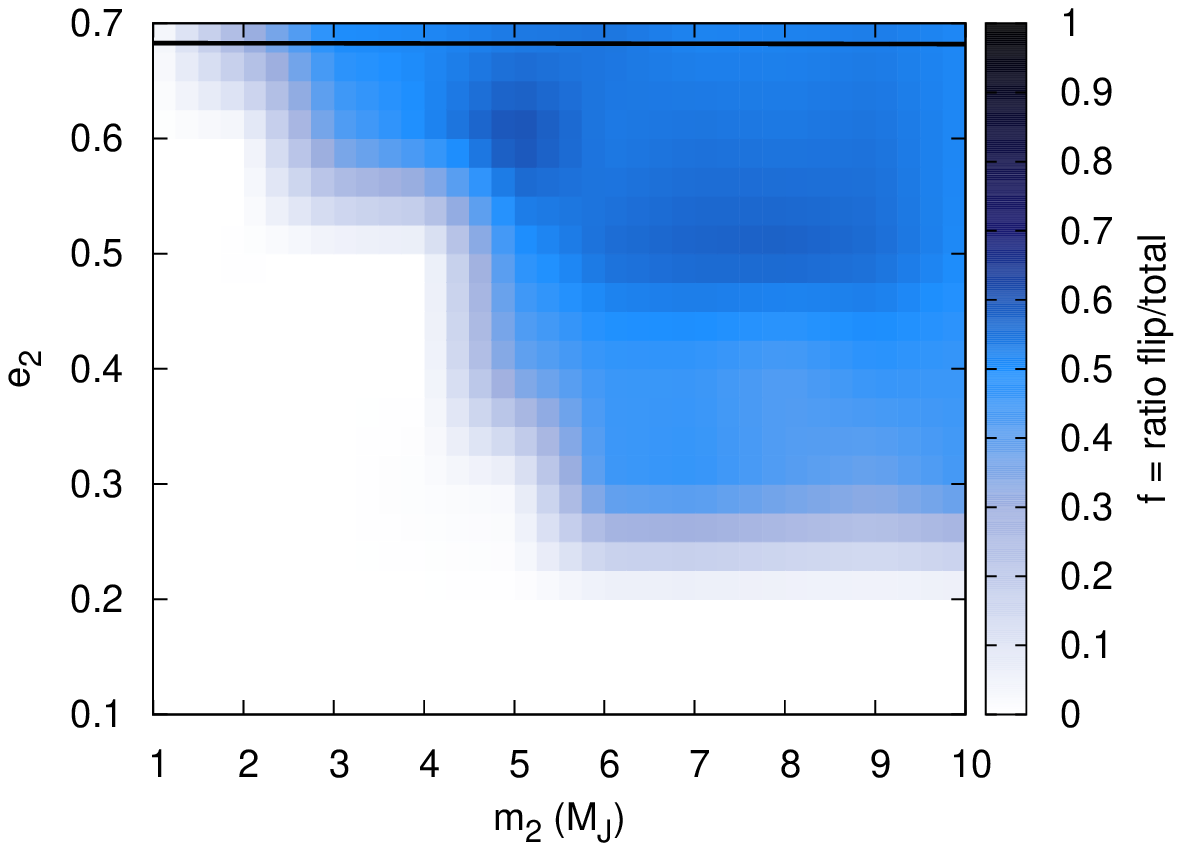}
      \caption{Constant parameters are $m_{1}=1\, \Mj$, $a_1=5$~AU, $a_2=61$~AU and $i_\tot=65^{\circ}$. Here $m_2$ varies from 1 to 10\, $\Mj$, and $e_2$ varies from 0.1 to 0.8. The black solid line marks the stability condition according to Equation (\ref{eq:Mar}). Above this line the system is unstable according to \citet{Mardling+01}. High eccentricities and massive perturbers  cause the inner planet to spend more time in retrograde orbits.}
      \label{map_e2m3_ratio}
   \end{minipage}\hfill   
\end{figure}

\begin{figure}[h!]
   \begin{minipage}[t]{0.40\linewidth}
      \centering \includegraphics[scale=0.65]{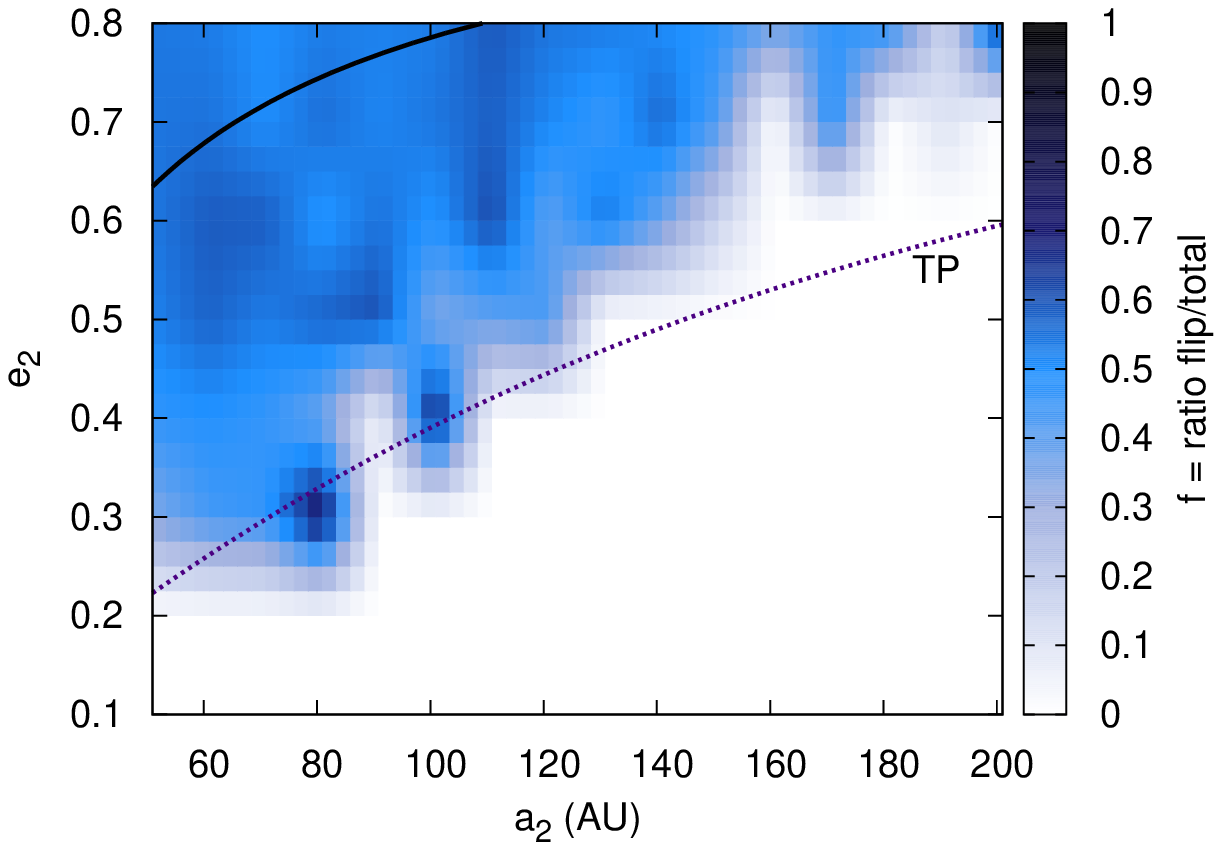}
      \caption{Constant parameters are $m_1 = 1\, \Mj$, $m_2 = 6\, \Mj$ and $i_\tot=65^{\circ}$. Here $a_2$ varies from 51 to 201 AU, and $e_2$ varies from 0.1 to 0.8. The black solid line marks 	the stability condition according to Equation (\ref{eq:Mar}). Above this  line the system is unstable according to \citet{Mardling+01}. The purple dotted line marks the flip criterion in the test particle limit. Systems above this line are expected to flip in the test particle limit. High outer eccentricities and a small SMA ratio cause the inner planet to spend more time in retrograde orbits.}
      \label{map_a2e2_ratio}
   \end{minipage}\hfill  
   \begin{minipage}[t]{0.48\linewidth}
      \centering \includegraphics[scale=0.65]{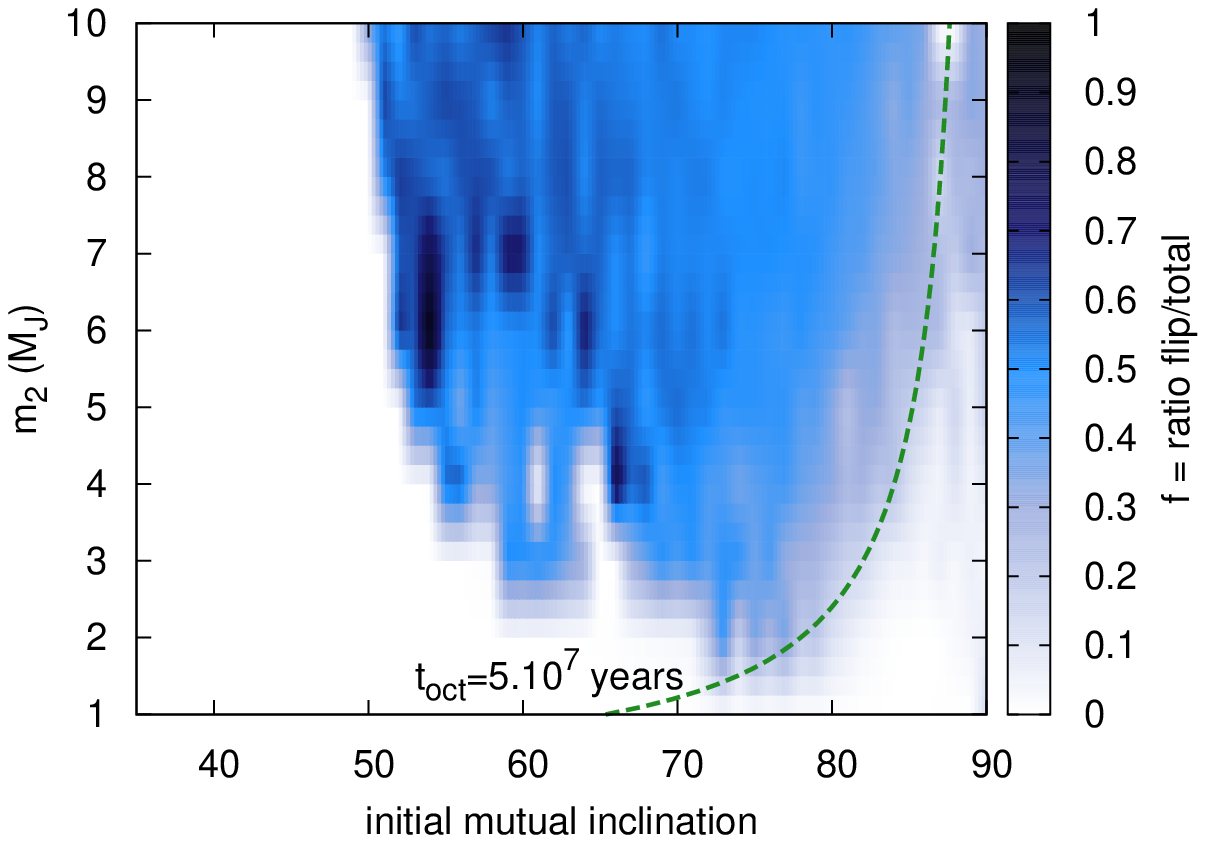}
      \caption{Constant parameters are $m_1=1\, \Mj$, $a_1=5$~AU, $a_2=61$~AU and $e_2=0.5$. The green dashed line gives the location where the octupole timescale is equal to $5\times 10^7$\, yr, for $e_1=0.9$. Mutual inclinations between $55^{\circ}$ and $85^{\circ}$ and perturber's masses between 4 and 10 $\Mj$ give the highest rate of retrograde configurations.}
      \label{map_im3_ratio}
   \end{minipage}\hfill
\end{figure}

\begin{figure}[h!]
   \begin{minipage}[t]{0.40\linewidth}
      \centering \includegraphics[scale=0.65]{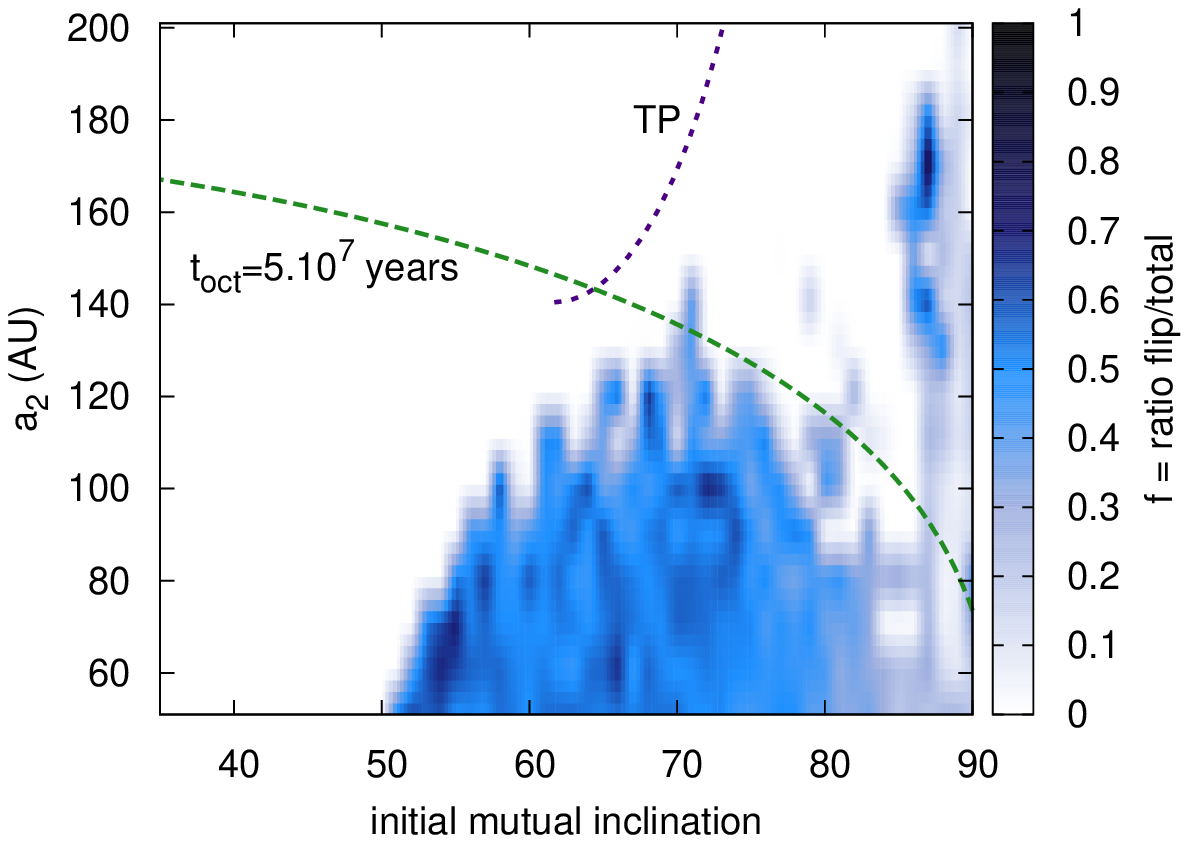}
      \caption{Constant parameters are $m_1 = 1\, \Mj$ and $m_2 = 6\, \Mj$ with $e_{2}=0.5$. The initial mutual inclination varies from $35^{\circ}$ to $90^{\circ}$, while $a_2$ varies from 51 to 201 AU. Systems beneath the purple dotted line are expected to flip in the test particle approximation. The green dashed line gives the location where the octupole timescale is equal to $5\times 10^7$\, yr, for $e_1=0.9$. To produce a retrograde orbit, the initial inclination must be in $[55^{\circ}:85^{\circ}]$ for most values of $a_2$.}
      \label{map_a2i_ratio}
   \end{minipage}\hfill  
   \begin{minipage}[t]{0.48\linewidth}   
      \centering \includegraphics[scale=0.65]{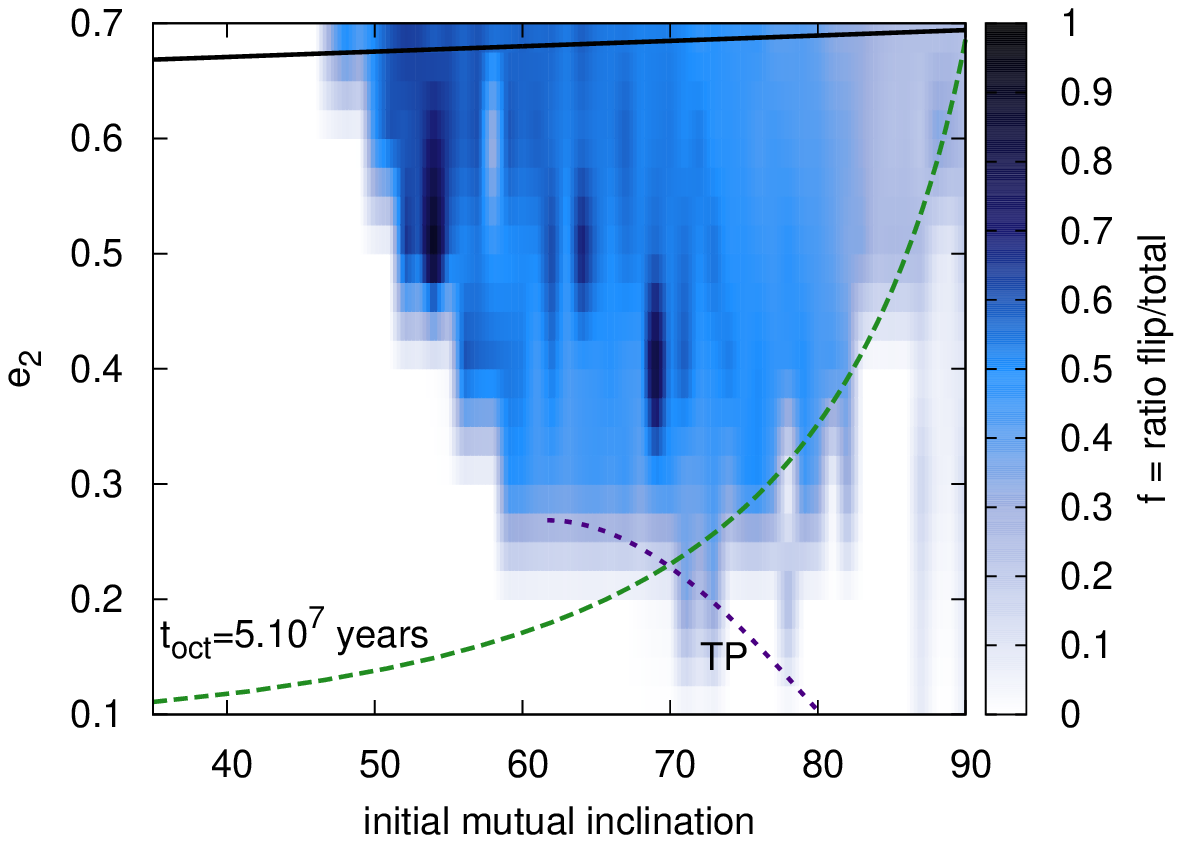}
      \caption{Constant parameters are  $m_{1}=1\, \Mj$, $m_2=6\, \Mj$, $a_{1}=5$~AU, $a_2=61$~AU. The top solid black line gives the stability limit: systems above this limit are likely to be unstable. Systems above the purple dotted line are expected to flip in the test particle approximation. The green dashed line gives the location where the octupole timescale is equal to $5\times 10^7$\, yr, for $e_1=0.9$. Highly eccentric and moderately high inclined companions cause the inner planet to spend more time in retrograde orbits.}
      \label{map_ie2_ratio}
   \end{minipage}\hfill
\end{figure}

\begin{figure}[h!]
   \begin{minipage}[t]{0.40\linewidth}
      \centering \includegraphics[scale=0.65]{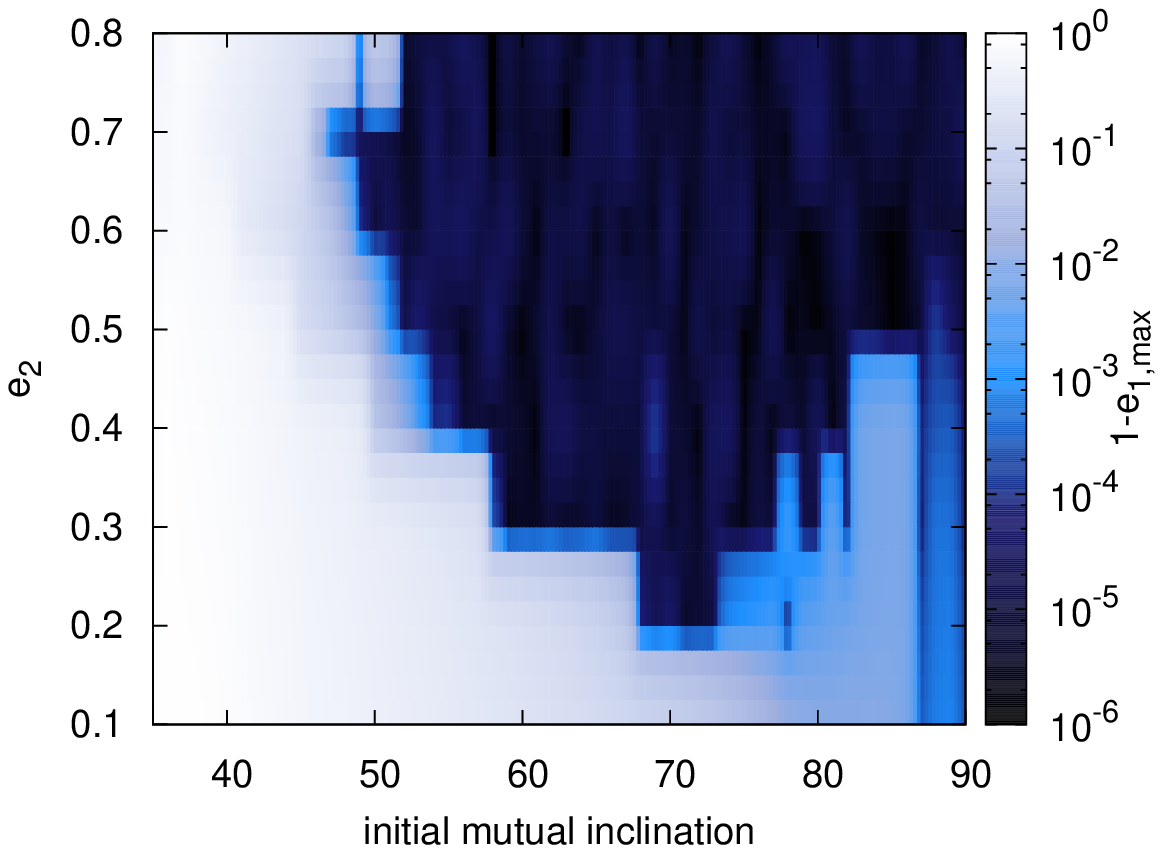}
      \caption{Maximum inner eccentricity (given as $1-e_{1,\rm max}$ in logarithmic scale) for the run in Figure \ref{map_ie2_ratio}. Very high eccentricities are associated with flips of the inner orbit.}
      \label{map_ie2_maxe1}
   \end{minipage}\hfill  
   \begin{minipage}[t]{0.48\linewidth}
      \centering \includegraphics[scale=0.65]{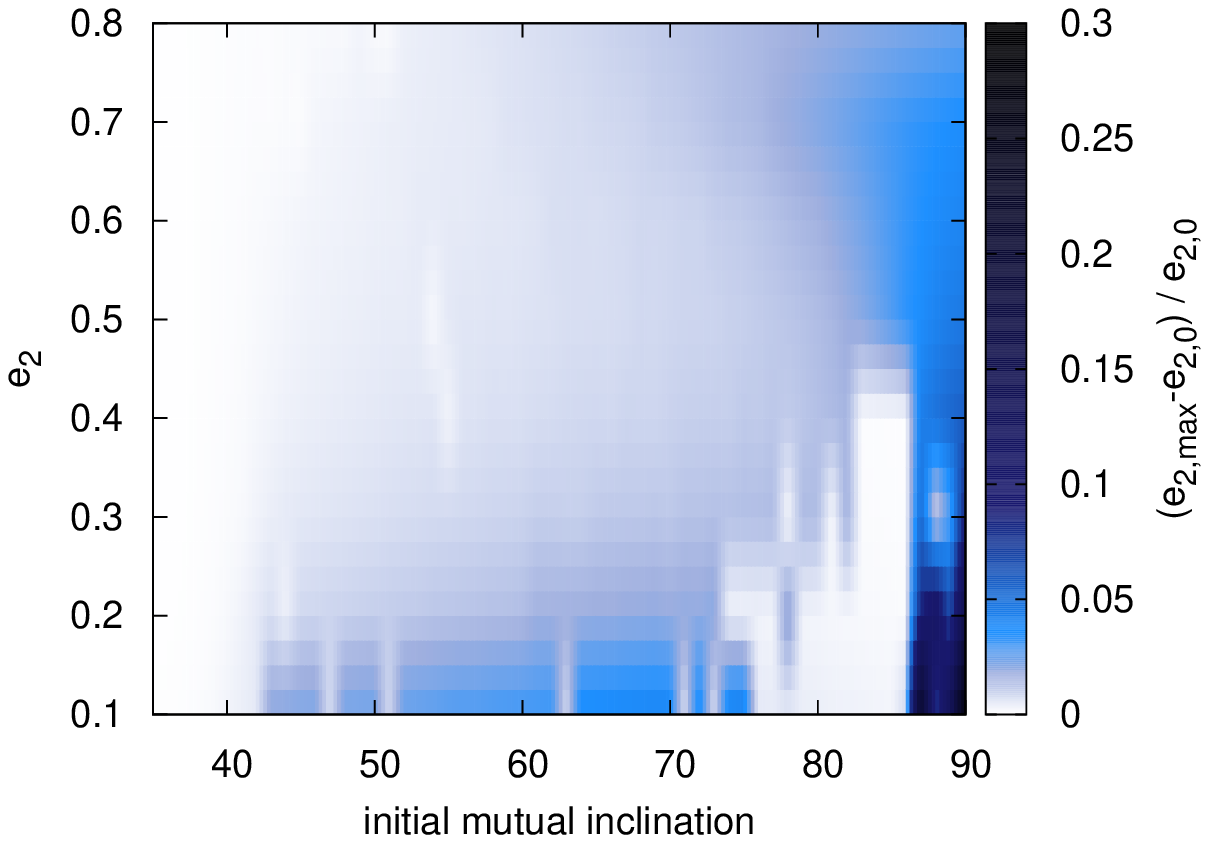}
      \caption{Variation of the outer eccentricity $e_2$ for the run in Figure \ref{map_ie2_ratio}. The color scale shows $(e_{2,\rm max}-e_{2,0})/e_{2,0}$, where $e_{2,0}$ is the initial outer eccentricity. This map indicates that the back reaction from the inner planet on the outer planet is more important at high mutual inclination and low eccentricities.}
      \label{map_ie2_maxe2}
   \end{minipage}\hfill
\end{figure}

\begin{itemize}
\item {\it Varying $m_1$ and $m_2$.}
In Figure \ref{map_m2m3_ratio} we find that as long as $m_2$ is at least twice as large as $m_1$, systems always have $f>0$. Furthermore, for $m_2\gtrsim 3m_1$, almost all the systems converge to $f\simeq 0.5$. A system with $m_2\leq 2 m_1$ will produce retrograde planets for a very limited zone in the phase space. On the contrary, a system with a more massive perturber, even if the latter is on a distant orbit, makes the inner body go into a retrograde motion over the course of 8 billion years. 

\item {\it Varying $a_1$ and $a_2$:}
In Figure \ref{map_a1a2_ratio} we show that for a large range of SMA, the value of $f$ only depends on the ratio between $a_1$ and $a_2$ (rather than the actual value of $a_1$ and $a_2$). 
This is, of course, not surprising because of the nature of the expansion. As mentioned above we shade in black the possible instability region according to Equation (\ref{eq:Mar}). With the parameters used for the runs of Figure \ref{map_a1a2_ratio}, we find that there are no more flips when $a_2\gtrsim 25\times a_1$. Note that this value could be different for other parameters.

\item {\it Varying $a_2$ and $m_2$.} Results in Figure \ref{map_a2m3_ratio} show that the  probability of reaching highly inclined orbits strongly depends on these two quantities. Strong outer perturbative potentials produce more flips of the inner orbit. We attribute the sharp transition between flips and no flips to the fact that the post-Newtonian timescale becomes dominant over the octupole timescale. In Figure  \ref{map_a2m3_ratio}, the green dashed line gives the approximate location for which the octupole timescale is equal to the 1PN timescale. Systems on the left-hand side of this line should not flip, as the post-Newtonian timescale becomes shorter than the octupole timescale. Note that this is just an approximate location, since the octupole timescale gives a rough evaluation for the behavior of the system. Furthermore, the octupole timescale is highly sensitive to the inner orbit eccentricity, which varies during the system evolution. Thus we show two possibilities, one with $e_1=0.9$ and one for $e_1=0.99$.

\item {\it Varying $e_2$.} 
The eccentricity of the perturber plays a significant part in the evolution of the inner orbit (see Figures \ref{map_e2m3_ratio}--\ref{map_a2i_ratio}). We find that the eccentricity of the perturber should be higher than $0.2$ at least in order to form retrograde inner planets.  This is an important constraint on the nature of these systems, and it is interesting to emphasize that a flip can be achieved already for nominally low eccentric perturbers such as $e_2=0.25$. We note that for our choice of fiducial parameters, perturbers with eccentricity higher than about 0.68 are unstable according to Equation (\ref{eq:Mar}). As shown in Figure \ref{map_ie2_ratio},  relaxing the test particle approximation yields a qualitatively different result. Specifically, in contrast to the test particle  case, the flip is, in fact, suppressed at large inclinations, and seems focused (for the nominal example) around initial inclinations $i_\tot\sim 70^{\circ}$. For comparison the test particle  flip criterion is depicted in these figures \citep[see][]{LN11,BK}. This criterion is symmetric around $90^\circ$, since the outer orbit remains fixed, and is valid only in the regime of very high inclinations \citep[$i_\tot>61.7^{\circ}$, see][]{BK}. Note that in Figure 10 the corresponding eccentricity for inclinations larger than $80^{\circ}$ falls below 0.1. If this criterion were valid, all planets above the line labeled ``TP'' in Figures \ref{map_a2e2_ratio} and \ref{map_ie2_ratio}, and below this line in Figure \ref{map_a2i_ratio}, should flip from prograde to retrograde, and because of the long integration time, converge to $f=0.5$.

\item {\it Varying $i_\tot$:}
The ratio of time spent on a retrograde orbit also depends on the initial mutual inclination of the system as seen in Figures \ref{map_im3_ratio}, \ref{map_a2i_ratio} and \ref{map_ie2_ratio}. For initial inclinations lower than $\sim 40^{\circ}$, since we set initially $e_1\to 0$, there are no strong excitations of the inclination and eccentricity, and therefore no possibility of flipping the orbit above $90^{\circ}$. On the other hand, we find that starting with a very highly inclined orbit ($i_\tot>85^{\circ}$) does not necessarily imply the formation of retrograde orbits. For most inclinations within the range $[55^{\circ},85^{\circ}]$, the evolution does not show a strong dependence on the initial inclination. This is one of the main differences with the test particle, where both the maximum inclination and eccentricity were well-defined functions of the initial inclination (see Section 2). In the TPQ, $f$ should remain equal to zero, since the initial inclination is lower than $90^{\circ}$. Thus it appears that an initial inclination between $55^{\circ}$ and $85^{\circ}$ is more likely to form retrograde planets. 
Concerning  the issue of why very high initial mutual inclinations ($i_\tot>85^{\circ}$) do not favor the production of retrograde orbits,  we show on Figures \ref{map_im3_ratio}, \ref{map_a2i_ratio} and \ref{map_ie2_ratio} the line at which $t_{\rm oct}=5\times 10^{7}$\, yr (taking $e_1=0.9$). This line strongly suggests that the suppression of the flip at high inclinations is because the octupole timescale becomes too large. Systems set initially with large inclinations typically have $t_{\rm oct}>5\times 10^{7}$\, years, which renders the triggering of the EKL mechanism less likely. In such systems, the inner planet does not enter high-eccentricity excitation phases, and is therefore less likely to end up as a HJ. Note that, of course, the inclination also changes as a function of time; however, the system will oscillate between large inclination (minimum eccentricity) and low inclination (large eccentricity). Considering $t_{\rm oct}$ for a large eccentricity means that the corresponding  inclination should be small, so we approximate it by the initial inclination.
 
\end{itemize}
  
We also show in Figures \ref{map_a2e2_ratio}, \ref{map_a2i_ratio} and \ref{map_ie2_ratio} the analytical prediction for a flip (depicted by purple dashed lines) using the derivation from \citet{BK}, which is valid only for inclinations larger than $61.7^{\circ}$. This presents the qualitatively different results between the  test particle approximation and our case. First we find that in our case, unlike the test particle  approximation, there is no symmetry of the flip condition around $90^\circ$, and in fact smaller inclinations (around $70^\circ$) are preferable. Furthermore, we can find occasions where a flip can happen in regions unreachable  in the test particle approximation, for example the low outer orbit eccentricity case with  $a_2<100$~AU as seen in Figure  \ref{map_a2e2_ratio}. It is important to note that the maximum quadrupole inner orbit's eccentricity did not shift from $90^\circ$ to $70^\circ$, however, the contribution of the octupole level of approximation yields a smaller probability for a flip at high inclinations. Furthermore, in the case of small mass of the perturbers the inner orbit  torques the outer orbit. This is more apparent in Figure \ref{map_im3_ratio} which shows that for larger masses we recover the test particle results. There are three ways to overcome torquing the outer orbit: first, by having a more massive perturber (as in the test particle approximation); second, by taking orbits with low initial mutual inclination (since the torque is proportional to $\sin i_\tot$ the torque is larger at high inclination), andn finally, by having a larger separation between the inner and outer orbits. The latter not only reduces the torque by reducing the length of the ``arm'' but also suppresses the octupole contribution. This behavior is apparent, for example, in Figure 9 where an ``island'' of large probability of flips appears at high inclination and large separations. Of course, large separations also reduce the octupole contribution, resulting in an isolated island.

As in the test particle case the system oscillates back and forth  from prograde to retrograde. However, unlike the test particle case, the system does not converge to  $f=0.5$, since the outer orbit reacts to the gravitational perturbations of the inner orbit. In fact one would expect that the system will prefer the retrograde motion since it is more stable \citep{Innanen}, which perhaps can explain the ``islands'' for which $f>0.5$. Note that we have tested in detail the convergence of our systems, and for our integration time ($8$\, Gyr) most of the runs already converged (see Appendix \ref{sec:conv}). Another interesting regime that arises from the parameter maps is a ``transition zone'' where the inner planet spends only about $10\%-20\%$ of its time on a retrograde orbit (colored pale blue in the figures).

Also important are the behaviors of the inner and outer orbits' eccentricities. In Figure \ref{map_ie2_maxe1} we show the maximum $e_1$ reached in the corresponding run of Figure \ref{map_ie2_ratio}, and in Figure \ref{map_ie2_maxe2} we show the (relative) maximum $e_2$ for the same run. Not surprisingly, the behavior closely resembles that of the test particle approximation. The  probability of flipping the orbits matches the maximum value of $e_1$: flips are associated with excursions to very high eccentricities, which, in fact, happen just before the flip. We find excursions of at least $1-e_{1,\rm max}\lesssim 10^{-4}$ when $f\simeq 0.5$.  Furthermore, in our case, the outer orbit's angular momentum is changing too, as can be seen in Figure \ref{map_ie2_maxe2} where we show the maximal relative value reached by the outer eccentricity. This plot shows that the suppression of flips at high initial mutual inclinations is highly related to the outer orbit's evolution. When the outer orbit's eccentricity almost does not change (marked in pale blue) the inner orbit is more likely to flip. 

These numerical results suggest that HJs that formed through planet--planet secular interactions should have a massive ($\geq 3\, \Mj$), eccentric ($\geq 0.25$), companion with a SMA between 50 and 100 AU, and a mutual inclination between 55$^{\circ}$ and 85$^{\circ}$. A planetary companion like this can drive a Jupiter-like planet in $5$~AU to a large eccentricity, which in the presence of dissipation can result in shrinking the orbit to form a HJr \citep[see][]{SN11a}.

In Appendix \ref{sec:max}, we study the distribution of another variable of interest, the maximum mutual inclination reached by the same systems as the ones studied in this section. We show that systems for which $f>0$ all reach the same maximum inclination of about $140^{\circ}$ which is one of the critical Kozai angles.


\subsection{Inner orbit eccentricity distribution}
\label{sec:cumul_dist}

\begin{figure}[h!]
      \centering \includegraphics[scale=1]{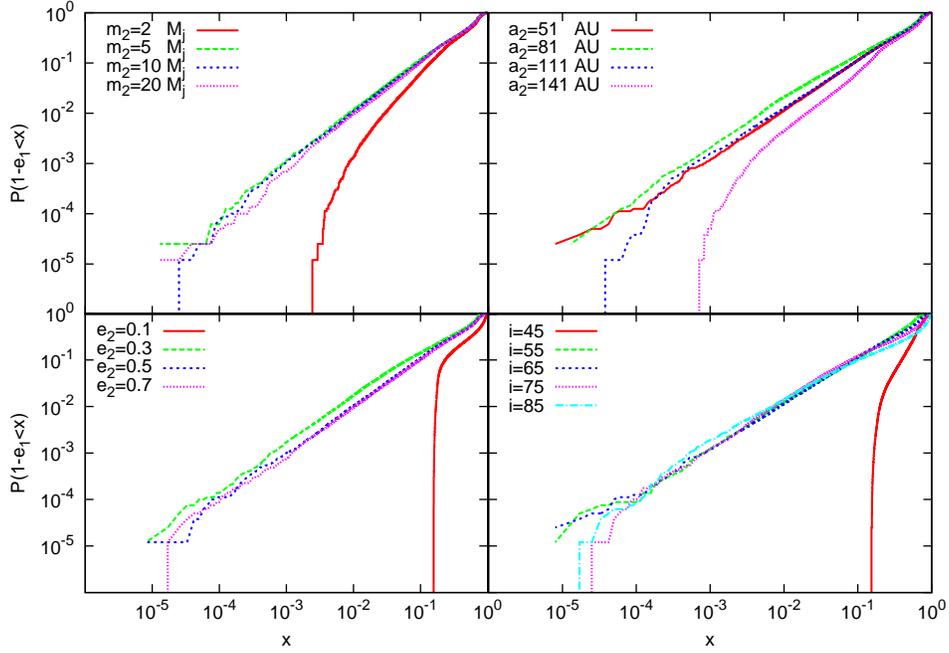}
      \caption{Cumulative distribution of the inner eccentricity (represented as $1-e_1$ in logarithmic scale) for the following system: a $1\, \Ms$ star with an inner planet of $1\, \Mj$ on an initially circular orbit at 5 AU. When not noted otherwise, the perturber has a mass of $6\, \Mj$ at 61 AU with an eccentricity of 0.5, and the two orbits are initially separated by $65^{\circ}$. For each panel we vary one of these parameters.
\textit{Top left:} we vary the mass of the perturber from 2 to 20 $\, \Mj$. \textit{Top right:} we vary the semi-major axis of the perturber from 51 to 141 AU. \textit{Bottom left:} we vary the eccentricity of the perturber from 0.1 to 0.7. \textit{Bottom right:} we vary the initial mutual inclination from $45^\circ$ to $85^\circ$. The inner orbit reaches high eccentricities (making orbital flips more likely to happen) for a large set of parameters, almost independently of the exact value of these parameters.}
      \label{cumul_distrib}
\end{figure}

As noted before, we focus on the dynamical evolution and neglect dissipation throughout the paper. But tidal dissipation will become important when the inner planet reaches very high eccentricities. Therefore in this section we focus specifically on the inner orbit's eccentricity distribution for these systems. In Figure \ref{cumul_distrib} we show the cumulative distribution of the inner orbit's eccentricity for different outer orbit configurations. Although a flip ($i_\tot>90^{\circ}$) happens when the inner orbit's eccentricity reaches a minimum, it also happens right after a large-eccentricity peak \citep[see][for discussion]{LN11,SN13}; thus the large-eccentricity peaks are a good proxy for a flip and vice versa \citep[it is certainly the case for the test particle scenario as shown in][ and we show here \text{that it remains true when this approximation breaks down}.]{Naoz+12bin}

As shown in Figure \ref{cumul_distrib}, a systematically low inner orbit eccentricity excitation is achieved for a combination of one or more of the following conditions for the outer orbit: low mass, low eccentricity,  large orbital separation and low mutual inclinations. However, for high mutual inclinations ($\gtrsim 50^\circ$), outer orbit eccentricities ($\gtrsim 0.25$) and massive perturber ($\gtrsim 5\, \Mj$)  the cumulative distribution is insensitive to the initial conditions. For these cases, as soon as the octupole effects are triggered, the inner eccentricity reaches extreme values ($e_1\gtrsim0.99)$.  As a consequence a counterplay may take place between the nearly radial orbit, which drives the planet to the star, and tidal dissipation, which can shrink and circularize the planet's orbit. As shown in \citet{SN11a}, a fairly high percentage of planets formed by this mechanism end up as hot Jupiters.

\section{Statistical estimation through a Monte Carlo experiment}
\label{sec:MC}

We explore the statistical properties of two representative scenarios of systems that are not only significantly different from the test particle approximation, but also distinct from one another. In the first scenario, we consider a perturber with a mass of 2~$\Mj$ (comparable to that of the inner planet, 1~$\Mj$), at $a_2=61$~AU. Such a system was shown in the previous section to suppress the EKL behavior. In the second scenario, we consider a system with a perturber with a mass of 6~$\Mj$  at $a_2=61$~AU. We showed that such a system can undergo large inclination and eccentricity oscillations, but still significantly differs from the test particle approximation since the EKL mechanism is suppressed near initial perpendicular configurations. As shown in Figure \ref{multi_conv} in Appendix \ref{sec:conv} most of these systems have converged after 2--6 Gyr. We run our integration up to 8 Gyr and study the distributions of the inclination as well as the inner and outer eccentricities for these systems at this arbitrary time of 8 Gyr, after they have reached a dynamical steady state.

\begin{figure}[h!]
      \centering \includegraphics[scale=0.7]{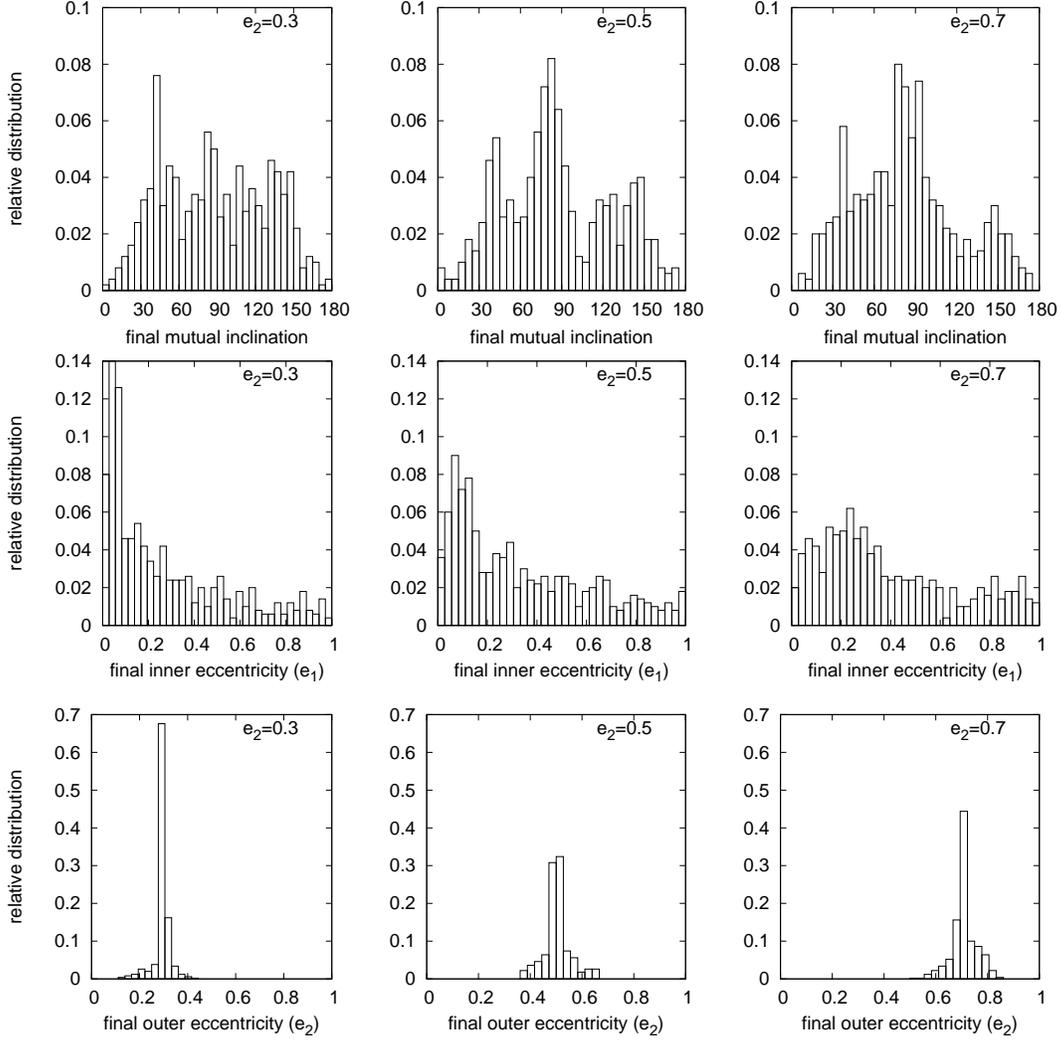}
      \caption{Final distribution of mutual inclination (top row), inner eccentricity (middle row) and outer eccentricity (bottom row) for a system with two planets. The inner one has $m_1=1\, \Mj$, $a_1=5$~AU and $e_1=0.01$ and the outer one has $m_2=2\, \Mj$, $a_1=61$~AU and three different initial eccentricities: 0.3 (left column), 0.5 (middle column) and 0.7 (right column). The initial mutual inclination is drawn from a distribution uniform in $\cos i_\tot$ between $0^{\circ}$ and $180^{\circ}$ via a Monte Carlo simulation. When the outer eccentricity is small, the final distribution of the orbital elements remains very close to its initial value. The back reaction from the inner orbit on the outer one is important and only a few systems flip. This is because the masses of the two planets are similar.}
      \label{MC_m3_2}
\end{figure}

\begin{figure}[t!]
      \centering \includegraphics[scale=0.7]{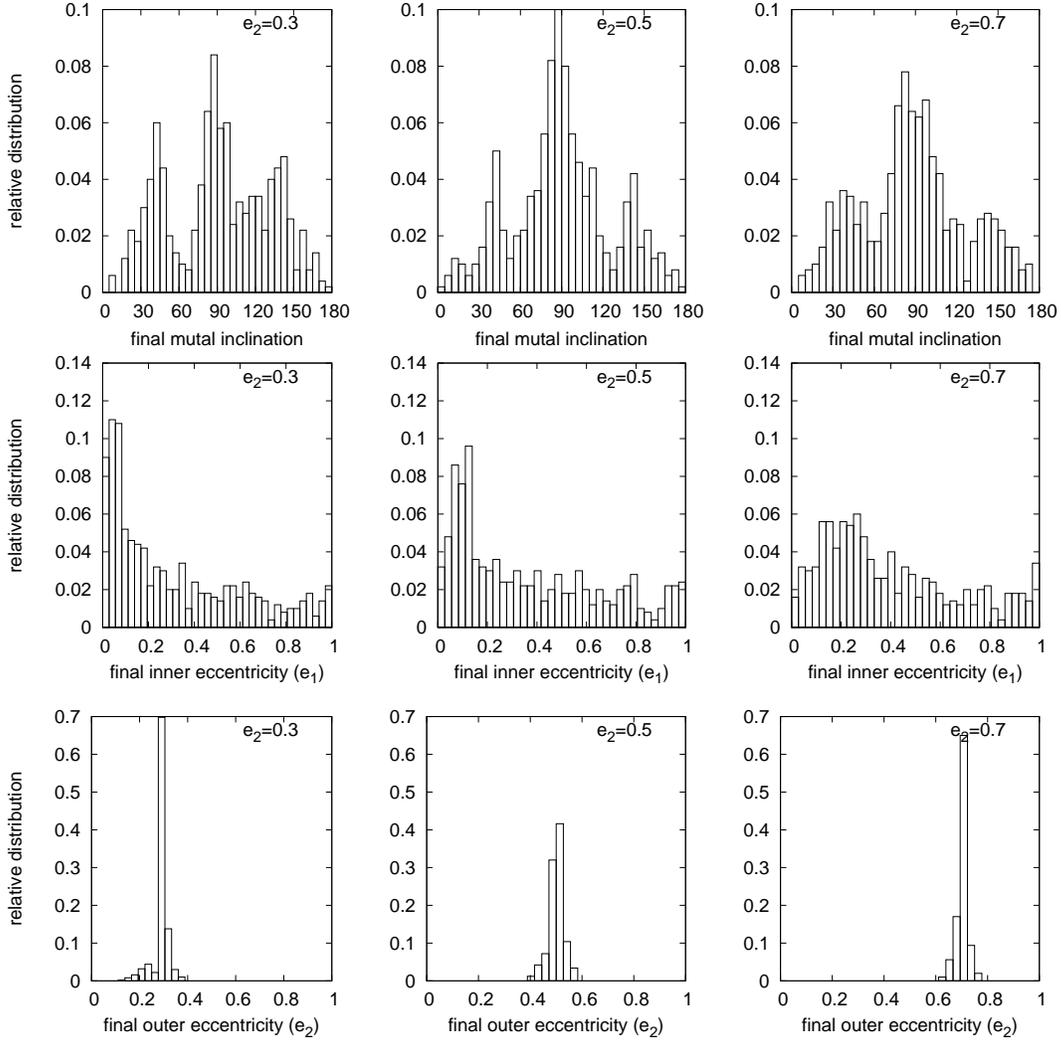}
      \caption{Final distribution of mutual inclination (top panels), inner eccentricity (middle panels) and outer eccentricity (bottom panel) for a system with two planets. The inner one has $m_1=1\, \Mj$, $a_1=5$~AU and $e_1=0.01$ and the outer one has $m_2=6\, \Mj$, $a_1=61$~AU and three different initial eccentricities: 0.3 (left column), 0.5 (middle column) and 0.7 (right column). The initial mutual inclination is drawn from a distribution uniform in $\cos i_\tot$ between $0^{\circ}$ and $180^{\circ}$ via a Monte Carlo simulation. The distribution of inclination is not uniform anymore and three peaks have appeared around $40^\circ$, $90^{\circ}$ and $140^\circ$. The inner eccentricity is well distributed between 0 and 1 and the outer one remains close to its original value. This is because of the larger mass of the outer planet, which limits the back reaction and favors orbital flips.}
      \label{MC_m3_6}
\end{figure}

For these runs we assume an isotropic distribution of the mutual inclinations (i.e., uniform in $\cos i_{\rm tot}$), between $0^\circ$ and $180^\circ$. We make a series of runs with initial conditions $m_1=1\, \Mj$, $a_1=5$~AU, $a_2=61$~AU, $e_1=0.01$ and $e_2=0.3,\ 0.5,\ 0.7$. We perform 500 runs for each set of parameters (3000 runs total). We show the results of these experiments in Figure \ref{MC_m3_2} for the case where $m_2=2\, \Mj$ and Figure \ref{MC_m3_6} for the case where $m_2=6\, \Mj$. Note that the choice of SMA allows the systems to achieve reasonable convergence (see appendix \ref{sec:conv}). As long as this convergence condition is fulfilled, the results should not be affected by our choice of SMA.

For all the cases with initial mutual inclination  lower than $40^\circ$ or above $140^\circ$, no large eccentricity and inclination oscillations occur, and there are no secular changes in the inclination and eccentricity. This  is because we set the inner orbit initially with nearly zero eccentricity \citep[we refer the reader to][for a discussion on the EKL behavior beyond the Kozai angles]{Li2013}. Interesting features appear when the system is initially in the Kozai regime. For the $m_2=2\, \Mj$  case  (Figure \ref{MC_m3_2}) with a small outer orbit's eccentricity, the EKL is not very efficient and the final distribution is similar to the initial one. The systems, however, still undergo ``classical'' quadrupole oscillations, during which they will always reach $40^\circ$ (provided they started on prograde orbits) or $140^\circ$ (provided they started in retrograde orbits) even if they do not flip. This results in the double peak (at $40^\circ$ and $140^\circ$) in the final inclination distribution \citep[as found in ][]{KCTF}, which is also the case for the  $m_2=6\, \Mj$ case (see Figure \ref{MC_m3_6}). As we enter the regime in the parameter space where the EKL mechanism begins to play a significant role (e.g., if the outer orbit's eccentricity is larger, or for a larger perturber's mass, as shown in Figure \ref{MC_m3_6}) we deviate from this classical behavior: an additional peak in the final mutual inclination appears around $90^\circ$ because the orbit now flips back and forth. The fraction of time that the orbits spend inclined at $90^\circ$, which is associated with minimum $e_1$, is larger than the fraction of time that the orbits spend at maximum eccentricity (minimum inclination). This, of course, also accounts for the large peak in the inner orbit's eccentricity near zero for the weak EKL case in Figures \ref{MC_m3_2} and  \ref{MC_m3_6}. As the EKL becomes more significant the minimum eccentricity shifts from zero and becomes wider. As expected, the outer orbit's eccentricity is sensitive to the outer planet's mass: for a massive outer planet, $e_2$ almost does not change, but for the  $m_2=2\, \Mj$ case, $e_2$ oscillates, which results in a suppression of the EKL mechanism.

%
%
%

\section{Summary and discussion}
\label{sec:discuss}

We have investigated numerically the dynamics of hierarchical triple systems consisting of a central star, an inner giant planet, and an outer, much more distant perturber that could be another giant planet or a substellar companion. We have varied systematically all important parameters including the separations and masses of the inner and outer planets, the mutual inclination, and the outer orbit's eccentricity (Figures \ref{map_a1a2_ratio} -- \ref{map_ie2_ratio}). We showed that relaxing the test particle approximation for this problem results in a much richer variety of dynamical outcomes. In contrast to the test particle case where extreme eccentricity peaks and flips of the inner orbit always happen around a mutual inclination of $90^\circ$, in our systems the behavior is quite different (e.g., Figures \ref{map_im3_ratio} and \ref{map_ie2_ratio}; see Section \ref{sec:ratio}) and the usual EKL behavior is confined to a smaller region of the parameter space. 

This study emphasizes the two interesting aspects of the EKL mechanism. One is the importance of the octupole level of approximation, which was explored in detail in the context of HJs and other astrophysical systems in \citet{SN11a,Naoz+12bin,SN13}. The second important aspect is related to relaxing the test particle approximation, which is shown to suppress the EKL mechanism for systems set initially close to perpendicular  configurations.  In this case, the outer orbit transfers some of its angular momentum to the inner orbit. We showed that the conditions for a flip to occur are sensitive to small changes in the outer orbit's angular momentum.

We have also shown that for a large set of parameters (most notably at large orbital separations and large mutual inclinations), the possibility of flipping the orbit from prograde to retrograde is suppressed when the octupole timescale becomes too long (Figures 8--10). Thus, for these systems the inner orbit eccentricity reaches smaller values (see Figure 11). If additional precession effects are present in the system, such as the post-Newtonian precession of pericenters, they can also become more important than the octupole timescale and overcome the octupole variations, suppressing the flips\footnote{If the 1PN timescale is shorter the  quadrupole timescale further eccentricity excitations are suppressed, when the two timescales are equal a resonant like behavior appears, as shown in \citet{Naoz+13GR}.}. The exact location of the limit between flip and no flip is approximated in Figure 5.
  
The Monte Carlo simulations we have conducted (Figures 13--14) suggest that statistically, a hierarchical triple   system that is far from the test particle approximation  is most likely to reside in three mutual inclinations regimes, near $\sim40^\circ$, $\sim90^\circ$ or $\sim140^\circ$, almost independently of initial conditions. Furthermore, initial conditions giving low eccentricities of the inner orbit are still more likely configurations to find such a system (although the more interesting behavior will happen in the large-eccentricity peaks). 

The parameter survey we have conducted gives strong and clear predictions for the orbital parameters of planetary companions that can result in the formation of a HJ. Figures \ref{map_a1a2_ratio}--\ref{map_ie2_ratio} suggest that the formation of HJs through planet--planet EKL mechanism predicts a massive ($>2\, \Mj$),  eccentric ($0.2-0.7$) planetary companion at separations of $\sim 50 - 140$~AU (for an inner Jupiter-like orbit set initially at $5$~AU). Furthermore, as expected for the Kozai mechanism a large ($>50^\circ$) mutual inclination is needed for an inner planet, set initially on a circular orbit. However, orbits close to perpendicular configurations are less likely to  flip the orbit (unlike the test particle approximation). This means that a planetary companion for misaligned HJs, most likely will not be perpendicular to the HJ orbit, but rather will have a mutual inclination in the range $55^\circ-85^\circ$.   

A planetary perturber such as described here should, in principle, cause very small variations of the radial velocity curve of the host star, in the form of a small linear trend. A planet of 6~$\Mj$ orbiting at 60 AU around a 1~$\Ms$ star with an eccentricity of 0.5, would cause a semiamplitude variation of 29 m s$^{-1}$ over its orbital period (465 years), which would appear as a linear acceleration of 0.12 m s$^{-1}$ year$^{-1}$. Such perturbations could only appear in long-term, high-precision radial velocity surveys. If we scale down the system slightly, and instead take an 8~$\Mj$ planet at 45 AU of a 0.5~$\Ms$ star with an eccentricity of 0.5, it would cause a semiamplitude variation of 63 m s$^{-1}$ over its orbital period (427 yr), which would appear as a linear acceleration of 0.3 m s$^{-1}$ yr$^{-1}$. Note that for theses calculations, we have assumed the orbital plane of the perturber to be aligned with the line of sight of the observer. If the angle between the two is $45^\circ$, the two trends previously calculated reduce to 0.08 m s$^{-1}$ yr$^{-1}$ and 0.21 m s$^{-1}$ yr$^{-1}$, respectively.

These predictions can also be used as a guide for future direct imaging observations such as the one presented in \citet{GPI06}, and can help differentiate between different perturbers (i.e., binary star or faraway planetary or brown dwarf companion).  An important caveat is that here we have studied two planet systems. Other routes to the formation of HJs in misaligned orbits exist, including interactions with stellar binary systems \citep[e.g.,][]{Naoz+12bin}, and even primordial misalignment of the disk with respect to the plane of the stellar binary \citep [e.g.,][]{Batygin12}. Achieving high eccentricity peaks (which may result in HJ formation) requires large initial mutual inclination (e.g., Figure \ref{cumul_distrib}) which can be a result of  planet--planet scattering \citep[e.g.,][]{CFMR}, dynamical relaxation \citep{PT2001} or early disk accretion from the surrounding gas envelope \citep{Thies+11}.


\section*{Acknowledgements}
We thank Guillaume H\'ebrard for useful discussions, and the anonymous reviewer for valuable comments and suggestions that improved the quality of the paper. Simulations for this project were performed on the HPC cluster \textit{fugu} funded by an NSF MRI award. This work was supported by NASA Grant NNX12AI86G at Northwestern University. SN was  supported by NASA through an Einstein Postdoctoral Fellowship awarded by the Chandra X-ray Center, which is operated by the Smithsonian Astrophysical Observatory for NASA under contract PF2-130096.



\bibliographystyle{apj}
\bibliography{biblio_HJ}


\appendix

\section{Appendix}

\subsection{Maximum inclination}
\label{sec:max}

Another variable well suited to map the dynamics of the system is the maximum mutual inclination $i_{max}$ reached by this system over one integration. Mapping the maximum inclination gives information about the dynamics: systems in which the EKL takes place should reach a maximum inclination of about $140^\circ$, whereas in systems where it does not happen, the maximum inclination should remain close to the initial one. In order to proceed to the complete mapping of the parameter space through $i_{\rm max}$, we successively vary the initial mutual inclination $i_\tot$, eccentricity $e_{2}$, SMA $a_2$ and mass $m_2$, in an iterative way, like in Section \ref{sec:ratio}. We plot the maximum mutual inclination between the two orbits as a function of two of these variables.

The results displayed here give additional information from the one presented in \S \ref{sec:ratio}. Mainly, when an orbit flips, it will always reach the critical Kozai angle of $140^{\circ}$. On the other hand, systems for which the EKL is negligible will have a maximum inclination close to its initial one. There is an intermediate zone, where the EKL is triggered but the exchange of angular momentum is not enough to flip the orbit. In this case, the maximum inclination reached by the system is located at $90^{\circ}$ (see, for instance, the light blue transition zone in Figure \ref{map_a2i_max}).

\begin{figure}[h!]
   \begin{minipage}[t]{0.40\linewidth}
      \centering \includegraphics[scale=0.65]{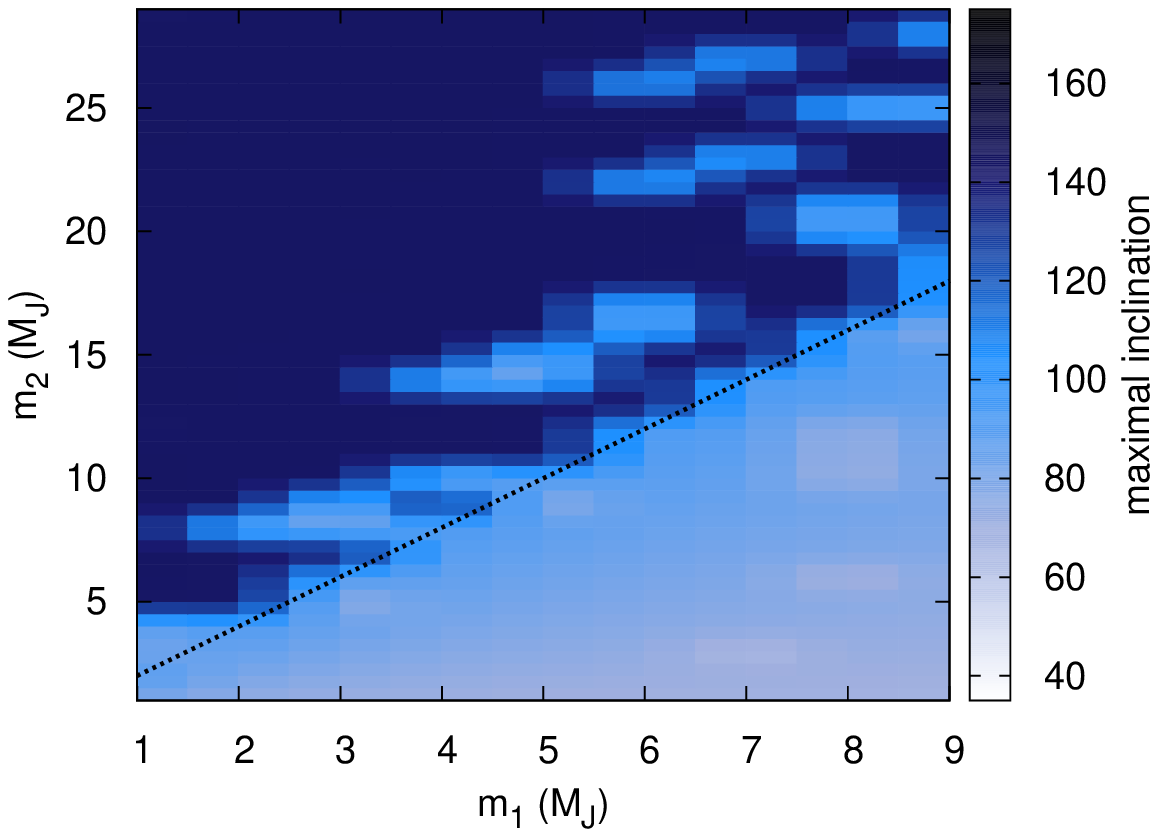}
      \caption{Constant parameters are  $e_{2}=0.5$, $i_\tot=65^{\circ}$, $a_1=5$~AU and $a_2=61$~AU. The black dashed line represents the $m_2=2m_1$ function.}
      \label{map_m2m3_max}
   \end{minipage}\hfill
   \begin{minipage}[t]{0.48\linewidth}
      \centering \includegraphics[scale=0.65]{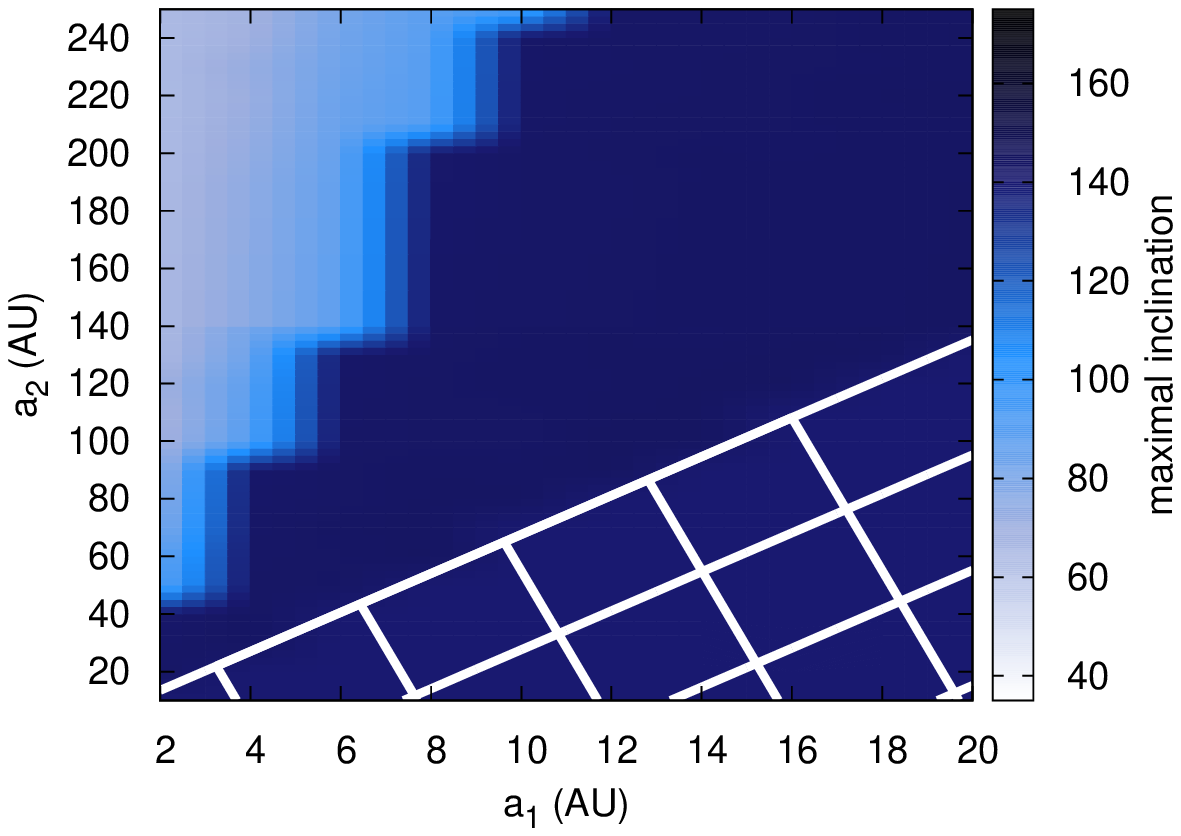}
      \caption{Constant parameters are $m_1=1\, \Mj$ and $m_2=6\, \Mj$, $e_{2}=0.5$ and $i_\tot=65^{\circ}$. The lower-right white-dashed region is where orbits are likely to be unstable (see \S 2).}
      \label{map_a1a2_max}
   \end{minipage}\hfill   
\end{figure}

\begin{figure}[h!]
   \begin{minipage}[t]{0.40\textwidth}
      \centering \includegraphics[scale=0.65]{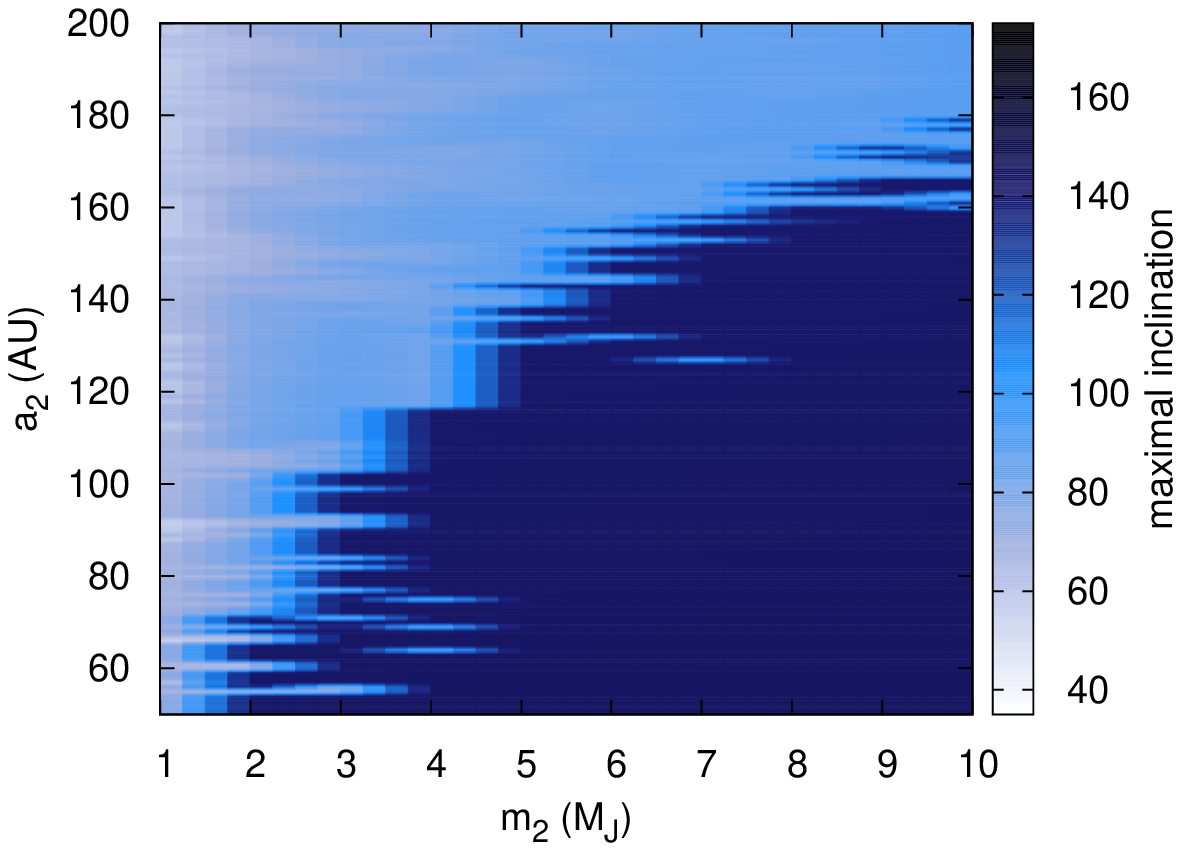}
      \caption{Constant parameters are $m_1=1\, \Mj$, $a_1=5$~AU, $e_2 = 0.6$ and $i_\tot=65^{\circ}$.}
      \label{map_a2m3_max}
   \end{minipage}\hfill
   \begin{minipage}[t]{0.48\linewidth}
      \centering \includegraphics[scale=0.65]{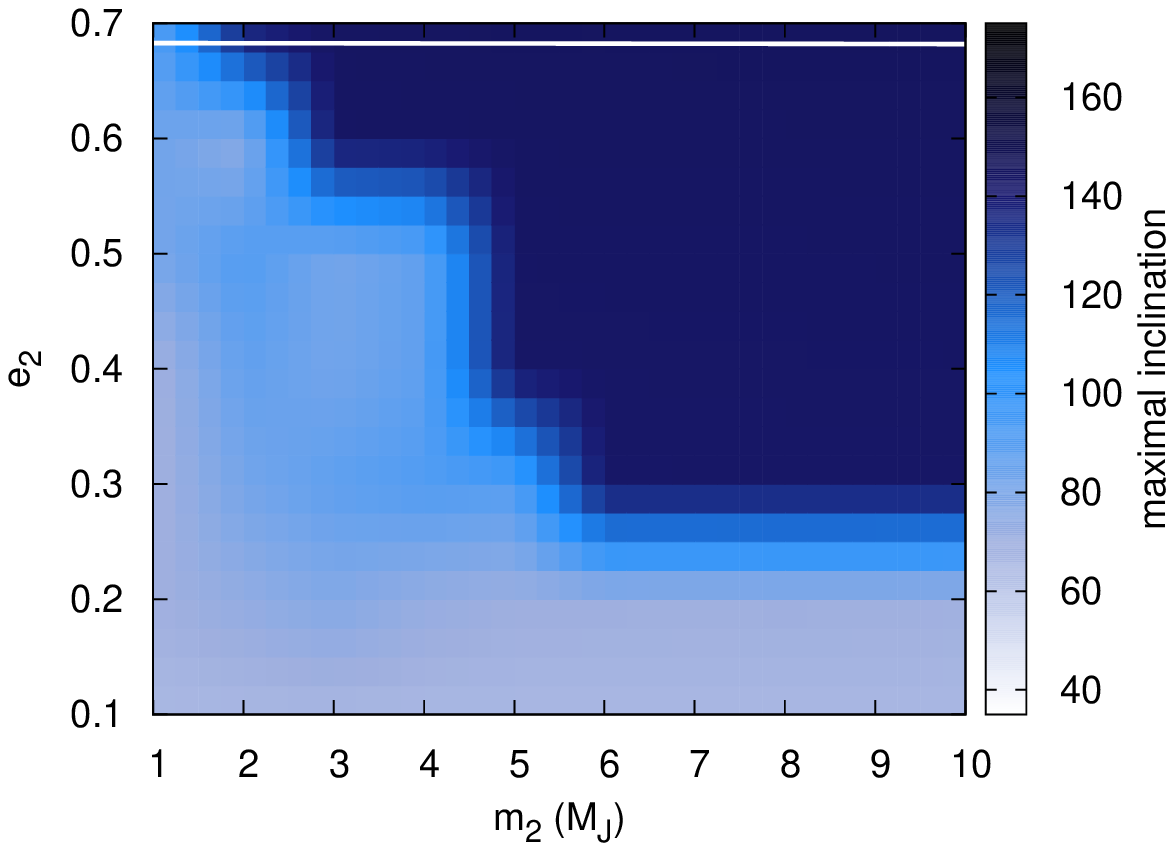}
      \caption{Constant parameters $m_{1}=1\, \Mj$, $a_1=5$~AU, $a_2=61$~AU and $i_\tot=65^{\circ}$.}
      \label{map_e2m3_max}
   \end{minipage}\hfill  
\end{figure}

\begin{figure}[h!]
   \begin{minipage}[t]{0.40\linewidth}
      \centering \includegraphics[scale=0.65]{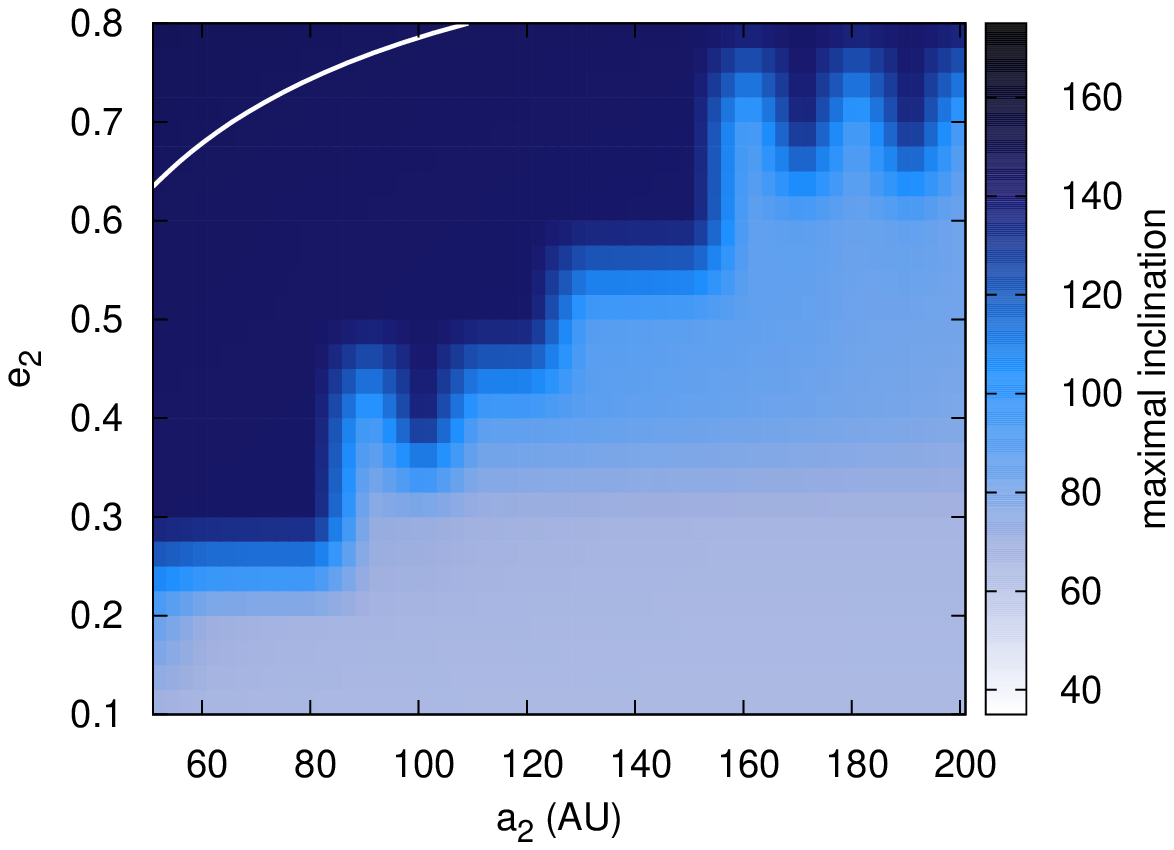}
      \caption{Constant parameters are $a_1=5$~AU, $m_1 = 1\, \Mj$ and $m_2 = 6\, \Mj$, with $i_\tot=65^{\circ}$.}
      \label{map_a2e2_max}
   \end{minipage}\hfill   
   \begin{minipage}[t]{0.48\linewidth}   
      \centering \includegraphics[scale=0.65]{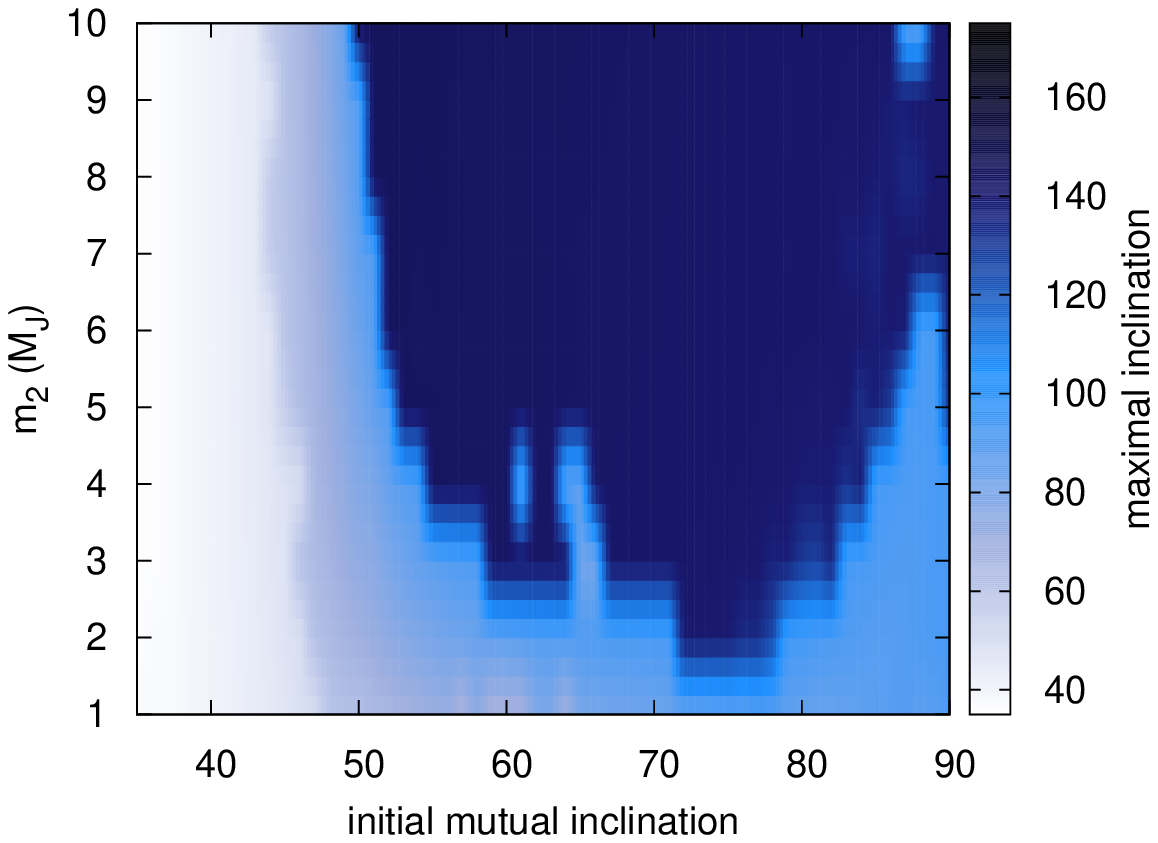}
      \caption{Constant parameters are $m_1=1\, \Mj$, $a_1=5$~AU, $a_2=61$~AU and $e_2=0.5$.}
      \label{map_im3_max}
   \end{minipage}\hfill
\end{figure}

\begin{figure}[h!]
   \begin{minipage}[t]{0.40\linewidth}
      \centering \includegraphics[scale=0.65]{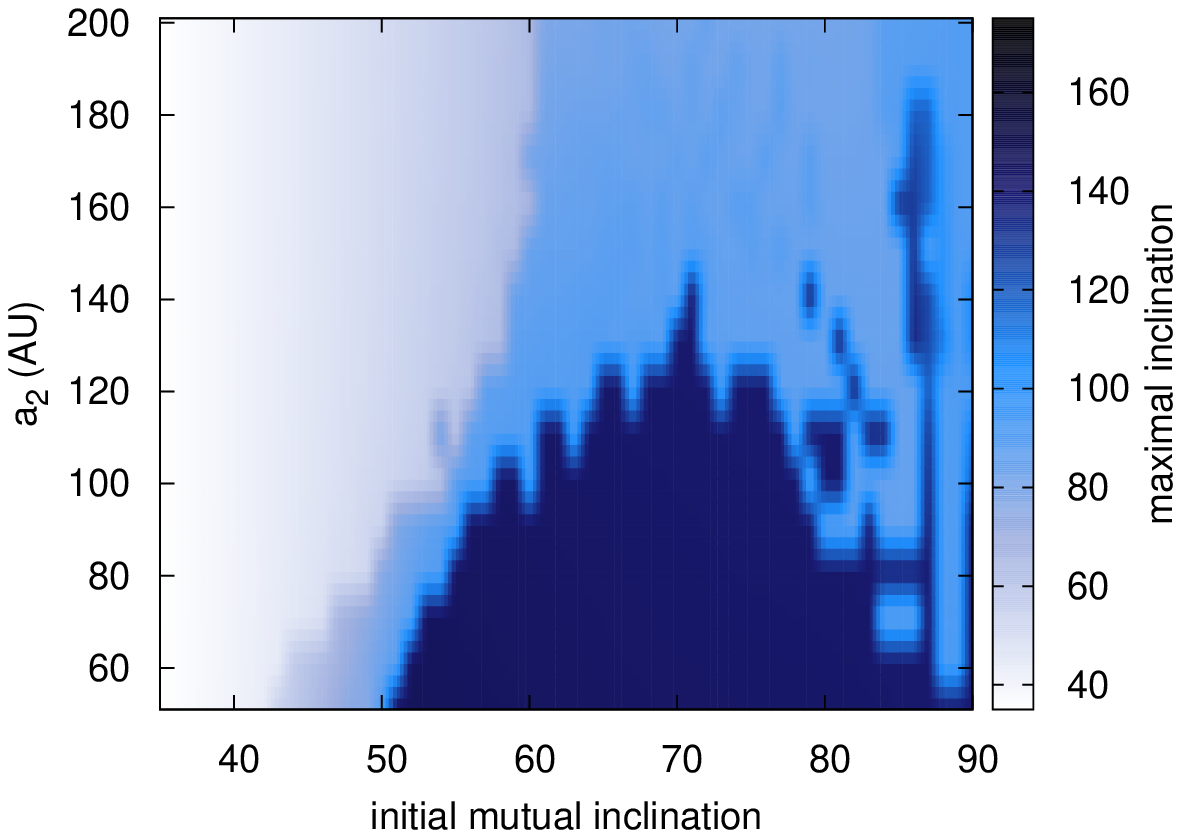}
      \caption{Constant parameters are $a_1=5$~AU, $m_1 = 1\, \Mj$ and $m_2 = 6\, \Mj$ with $e_{2}=0.5$.}
      \label{map_a2i_max}
   \end{minipage}\hfill
   \begin{minipage}[t]{0.48\linewidth}
      \centering \includegraphics[scale=0.65]{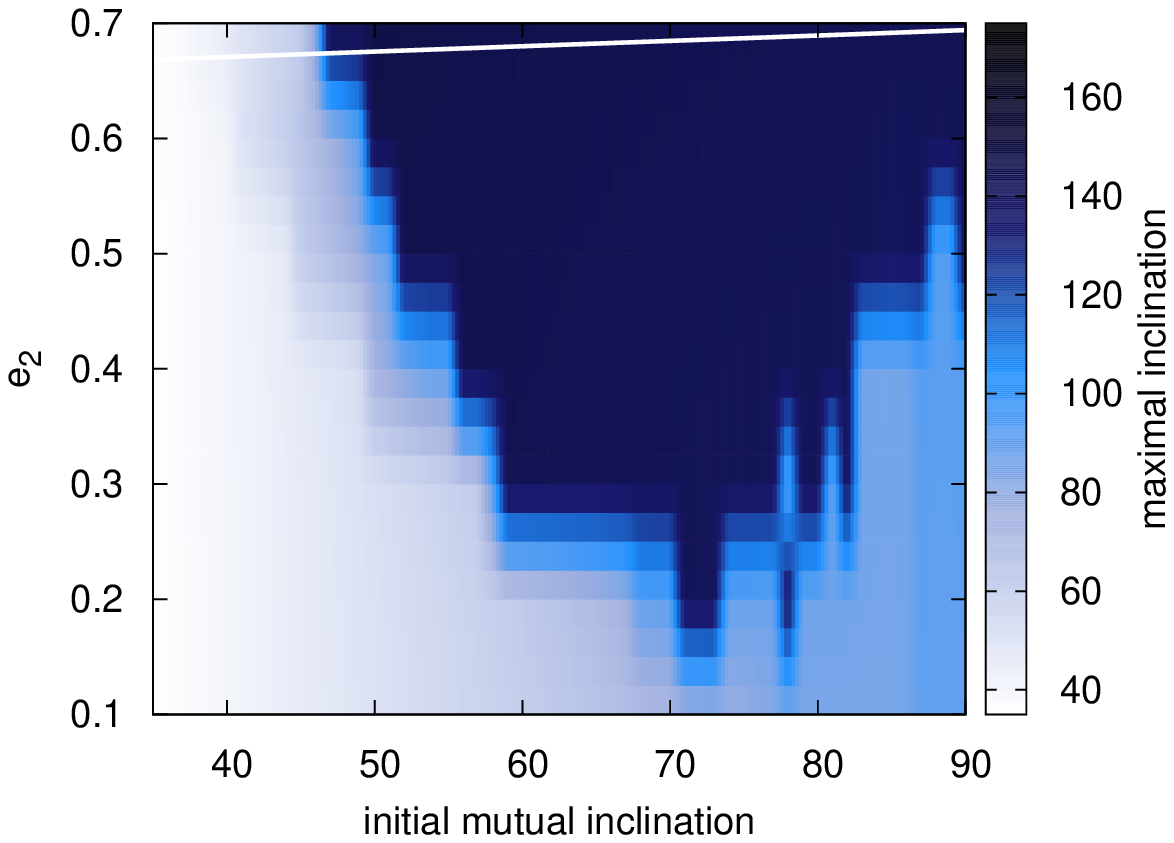}
      \caption{Constant parameters are $m_{1}=1\, \Mj$, $m_2=6\, \Mj$, $a_{1}=5$~AU, $a_2=61$~AU.} 
      \label{map_ie2_max}
   \end{minipage}\hfill
\end{figure}


\subsection{Convergence}
\label{sec:conv}

\begin{figure}[h!]
      \centering \includegraphics[scale=1]{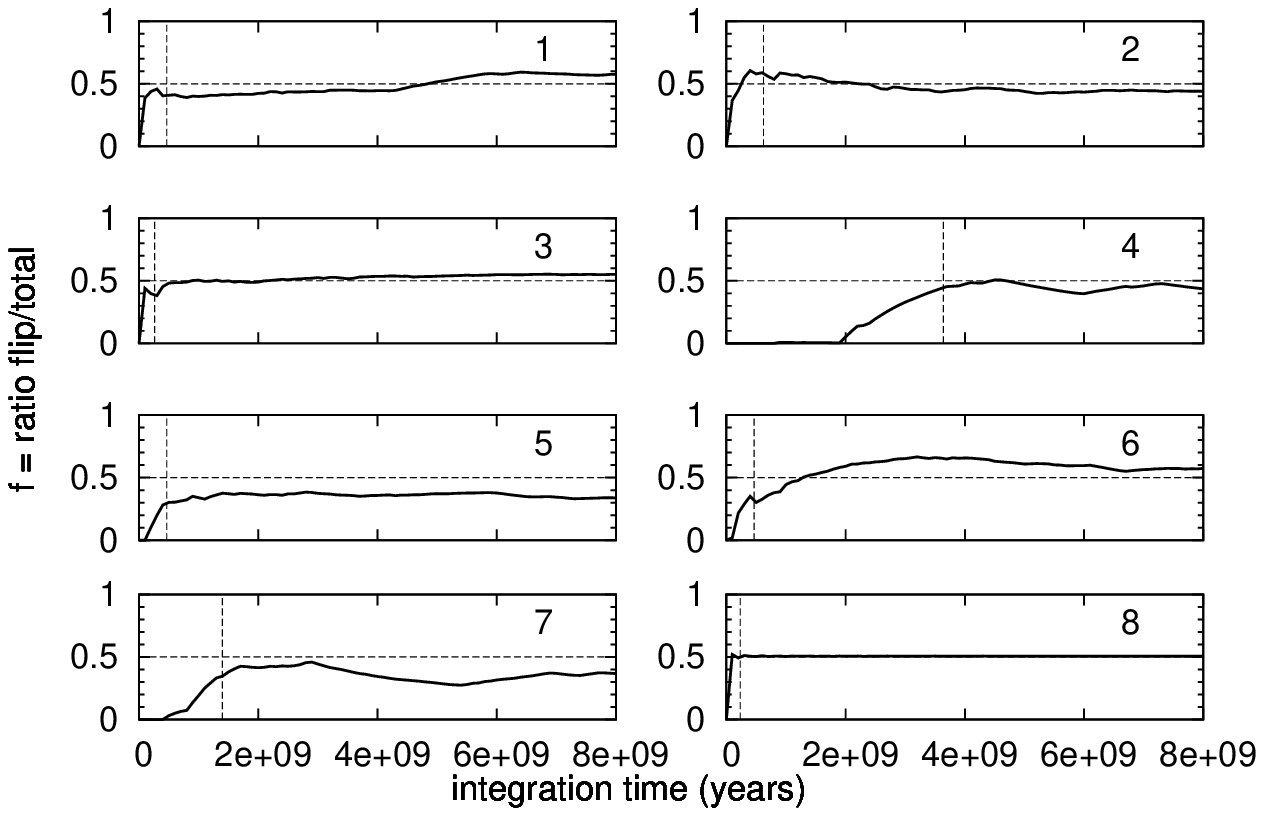}
      \caption{Value of $f$ as a function of the time of integration. The initial parameters of each run are given in Table \ref{tab:conv}, with the label of each run given in the top right corner of each panel. The horizontal dashed line indicates $f=0.5$ and the vertical dashed line indicated $t=1000\times t_{\rm quad}$.} 
      \label{multi_conv}
\end{figure}

We discuss the validity of choosing an integration time of 8 billion years. First, the age of the planetary systems provides an obvious physical limit. The oldest known star is 13.2 billion years old \citep{Frebel+07}. Conducting a large set of simulations is computationally expensive, which gives another limitation. We define a system as convergent for a given time of integration if $f$ has reached a constant value over this time. In Figure \ref{multi_conv} we show that an integration time of 8 billion years is sufficient for most systems to reach a steady state in $f$. Depending on the initial conditions (given in Table \ref{tab:conv}), $f$ takes a different time to reach a steady-state value. In the better case this value is achieved within less than a billion years, whereas when the perturbation is weak, it takes several billion years to converge. In order to use a timescale more relevant to each system, we use the period of Kozai oscillations, as given by Equation (\ref{eq:tquad}), where $P_1$ is the period of the inner planet. In Table \ref{tab:conv} we give the period of the Kozai oscillations for each system that we study. In all the runs, we indicate by a vertical line the time $t=1000\times t_{\rm quad}$. In some runs this timescale seems relevant to achieve convergence, whereas it fails in some others (see e.g., runs 1, 6 or 7). More precisely, the convergence is slower in the case of extreme initial conditions, such as low initial inclinations or large mass or semi-major axis ratios. An integration time between $5000\times t_{\rm quad}$ and $10000\times t_{\rm quad}$ appears safer in order to achieve convergence in all cases. It would, however, be numerically demanding (see ,e.g., run 4, where $5000\times t_{\rm quad}\simeq 20$ billion years), so we choose to restrain ourselves to 8 billion years. 

\begin{deluxetable}{lcccccc}
\label{tab:conv}
\tablewidth{300pt}
\tablecaption{Initial conditions of Figure \ref{multi_conv}}
\tablenum{2}
\tablehead{\colhead{Run} & \colhead{$m_2$} & \colhead{$a_2$} & \colhead{$e_2$} & \colhead{$i_\tot$} & \colhead{$f_{\rm final}$} & \colhead{$t_{\rm quad}$} \\ 
\colhead{} & \colhead{($\Mj$)} & \colhead{(AU)} & \colhead{} & \colhead{(deg)} & \colhead{} & \colhead{($\times 10^6$~yr)} } 
\startdata
1 & 6 & 61 & 0.5 & 65 & 0.576 & 0.466 \\
2 & 6 & 61 & 0.3 & 65 & 0.440 & 0.623 \\
3 & 6 & 61 & 0.7 & 65 & 0.551 & 0.261 \\
4 & 6 & 121 & 0.5 & 65 & 0.436 & 3.639 \\
5 & 6 & 61 & 0.5 & 80 & 0.339 & 0.466 \\
6 & 6 & 61 & 0.5 & 55 & 0.574 & 0.466 \\
7 & 2 & 61 & 0.5 & 75 & 0.368 & 1.399 \\
8 & 12 & 61 & 0.5 & 65 & 0.507 & 0.233 \\
\enddata
\tablecomments{Initial conditions of Figure \ref{multi_conv}. For all these runs, we took $m_0=1\, \Ms$, $m_1=1\, \Mj$; $a_1=5$~AU and $e_1=0.01$. $f_{\rm final}$ is the fraction of time spent on a retrograde orbit after 8 billion years, and $t_{\rm quad}$ is the period of Kozai oscillations calculated from equation (\ref{eq:tquad}).}
\end{deluxetable}

\end{document}